\documentclass{article}

\usepackage{PRIMEarxiv}

\usepackage[utf8]{inputenc} 
\usepackage[T1]{fontenc}    
\usepackage{hyperref}       
\usepackage{url}            
\usepackage{booktabs}       
\usepackage{amsfonts}       
\usepackage{nicefrac}       
\usepackage{microtype}      
\usepackage{lipsum}
\usepackage{fancyhdr}       
\usepackage{graphicx}       
\usepackage{caption}
\usepackage{tabulary} 
\usepackage[flushleft]{threeparttable}
\usepackage{multirow}
\usepackage{adjustbox}
\graphicspath{{media/}}     
\usepackage{subcaption}
\usepackage{amsmath,amsfonts,bm}

\usepackage[colorinlistoftodos,prependcaption,textsize=tiny]{todonotes}

\newcommand{\wf}[1]{{\color{black}{#1}}}

\newcommand{\tabincell}[2]{\begin{tabular}{@{}#1@{}}#2\end{tabular}}

\title{HelixFold-Single: \\MSA-free Protein Structure Prediction \\by Using Protein Language Model as an Alternative}

\author{
 Xiaomin Fang$^1$\thanks{Equal contribution.}, Fan Wang$^1$\footnotemark[1]  \thanks{Corresponding to wang.fan@baidu.com and songle@biomap.com.}, Lihang Liu$^1$\footnotemark[1], Jingzhou He$^1$,\\
 \textbf{Dayong Lin$^1$, Yingfei Xiang$^1$, Xiaonan Zhang$^1$, Hua Wu$^1$, Hui Li$^2$, Le Song$^2$\footnotemark[2]}, \\
 $^1$Baidu Inc., $^2$BioMap, \\
 }

\begin{document}
\maketitle

\begin{abstract}
AI-based protein structure prediction pipelines, such as AlphaFold2, have achieved near-experimental accuracy. These advanced pipelines mainly rely on Multiple Sequence Alignments (MSAs) as inputs to learn the co-evolution information from the homologous sequences. Nonetheless, searching MSAs from protein databases is time-consuming, usually taking dozens of minutes. Consequently, we attempt to explore the limits of fast protein structure prediction by using only primary sequences of proteins. HelixFold-Single is proposed to combine a large-scale protein language model with the superior geometric learning capability of AlphaFold2. Our proposed method, HelixFold-Single, first pre-trains a large-scale protein language model (PLM) with thousands of millions of primary sequences utilizing the self-supervised learning paradigm, which will be used as an alternative to MSAs for learning the co-evolution information. Then, by combining the pre-trained PLM and the essential components of AlphaFold2, we obtain an end-to-end differentiable model to predict the 3D coordinates of atoms from only the primary sequence. HelixFold-Single is validated in datasets \emph{CASP14} and \emph{CAMEO}, achieving competitive accuracy with the MSA-based methods on the targets with large homologous families. Furthermore, HelixFold-Single consumes much less time than the mainstream pipelines for protein structure prediction, demonstrating its potential in tasks requiring many predictions. The code of HelixFold-Single is available at
\url{https://github.com/PaddlePaddle/PaddleHelix/tree/dev/apps/protein_folding/helixfold-single}, and we also provide stable web services on \url{https://paddlehelix.baidu.com/app/drug/protein-single/forecast}.

\end{abstract}

\keywords{Protein structure prediction \and Primary sequence \and Protein language model \and Large-scale}





\section{Introduction}
Proteins participate in essentially all biological processes and play critical roles for an organism. The structures of proteins are highly correlated to their functions in the biological processes. Determining the protein structures to understand their functions can bring considerable contributions to life science.

In recent years, AI-based protein structure prediction technologies have made significant progress in prediction accuracy, demonstrating great prospects for the drug and vaccine industry. Particularly, AlphaFold2 \cite{jumper2021highly} pushes the performance to a new frontier in the challenging 14th Critical Assessment of protein Structure Prediction (CASP14 \cite{moult2005decade}), approaching the accuracy of experimental determination methods. Mainstream protein structure prediction pipelines heavily rely on co-evolution information extracted from Multiple Sequence Alignments (MSAs). MSAs can be simply regarded as protein chains similar to the target protein chain in sequence. MSA is related to the co-evolution information of protein sequences, which is crucial to predicting its structure. However, over-reliance on MSAs becomes the bottleneck of various protein-related tasks.

First, compared with the time (usually several seconds) required for model inference in the structure prediction pipeline, searching MSAs is time-consuming, costing dozens of minutes for a protein. The time-consuming searching is devastating in the tasks demanding high-throughput requests, such as protein design.
Second, the primary structures (single sequence), rather than the MSAs, drive the folding of the proteins. The MSA extracting methods are also not designed specifically for protein folding. Thus, the MSA-based pipelines only memorize the determined structures of similar proteins for prediction but do not entirely understand the mechanism of protein folding.


Consequently, designing an accurate MSA-free protein structure prediction method to address the mentioned issues is likely to benefit and accelerate the development of protein studies. We argue that a large-scale protein language model (PLM) can be served as an alternative to the MSAs to learn the co-evolution knowledge for MSA-free prediction. We speculate that a PLM with billions of parameters can effectively memorize the MSAs and infer the co-evolution information. The past few years have seen the tremendous success of large-scale language models \cite{vaswani2017attention,kenton2019bert, brown2020language} in Natural Language Processing, a field that shares a lot of characters with protein studying. With the increase of the model parameters, the capacity for learning language knowledge grows substantially. Using self-supervised learning on large-scale unlabeled proteins, PLMs can reveal the long-range relation along protein sequences and improve downstream protein-related tasks.
Advanced works have attempted to adopt PLMs to enhance the performance of multiple downstream tasks, such as estimating the secondary structures and the functions \cite{rao2019evaluating, elnaggar2020prottrans, rao2020transformer, xiao2021modeling}. Particularly, several studies \cite{chowdhury2021single, weissenow2022protein, wang2022single} attempted to apply PLMs to protein structure prediction. Most works first predict the inter-residue 2D geometry by neural networks and then reconstruct the 3D structure based on energy minimization, which can not provide end-to-end 3D structure prediction. Besides, compared with the geometric learning capability of EvoFormer and Structure Module proposed by AlphaFold, the capacities of the geometric models used by these methods, such as recursive models and ResNets, are also unsatisfactory in understanding the co-evolution and spatial relations between the residues in a single sequence.

Inspired by the progress of PLMs and AlphaFold2, we propose an end-to-end MSA-free protein structure prediction pipeline, HelixFold-Single. The model used in HelixFold-Single consists of two major components: a large-scale PLM as the foundation and the essential components from AlphaFold2 for folding. The PLM can encode the primary structure into single representation and pair representation to learn the domain knowledge. The EvoFormer and Structure Module from AlphaFold2 are then integrated to process the representation, learn the geometric knowledge, and then predict the coordinates of atoms. The two components are wired up to give an end-to-end differentiable model.
HelixFold-Single contains two training stages. In the first stage, the large-scale PLM is trained with thousands of millions of unlabeled single sequences by the task of masked language prediction. In the second stage, we train the whole model with the protein structures composed of experimental ground-truth and augmentation structures generated by AlphaFold2.

We compare HelixFold-Single with AlphaFold2 and RoseTTAFold on datasets \emph{CASP14} and \emph{CAMEO}. HelixFold-Single achieves competitive accuracy with those methods on proteins with sufficient homologous sequences. We also analyze the performance of HelixFold-Single on targets with various homologous sequences, and HelixFold-Single is capable of providing accurate structure predictions on most targets, especially the targets with large homologous families. The ablation study comparing the PLMs of different sizes demonstrates the importance of the size of PLM for structure prediction. Furthermore, HelixFold-Single shows great superiority in prediction efficiency compared with the MSA-based methods and could be applied to protein-related tasks demanding a great number of predictions. The code of HelixFold-Single is publicly released at GitHub \url{https://github.com/PaddlePaddle/PaddleHelix/tree/dev/apps/protein_folding/helixfold-single}. Web service of HelixFold-Single is also available at \url{https://paddlehelix.baidu.com/app/drug/protein-single/forecast} to provide efficient protein structure predictions.

\section{HelixFold-Single}
\begin{figure*}[t]
\centering
\includegraphics[width=1.0\columnwidth]{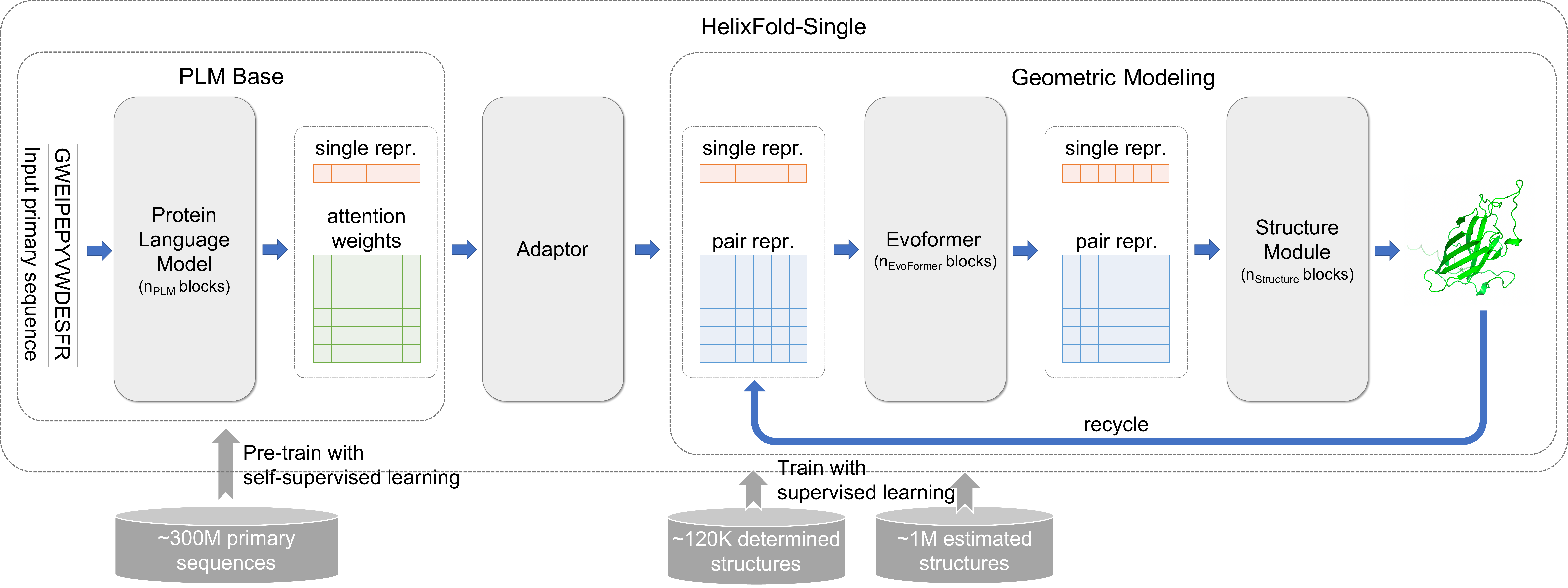}
\caption{The framework of HelixFold-Single with a protein language model as PLM Base, the compose of EvoFormer and Structure Module of AlphaFold2 as Geometric Modeling, and Adaptor to connect PLM Base and Geometric Modeling.}
\label{fig:framework}
\end{figure*}

HelixFold-Single aims to take advantage of both the protein language model (PLM) and the main modules used in AlphaFold2 for single sequence-based protein structure prediction. As exhibited in Figure~\ref{fig:framework}, HelixFold-Single consists of three components: PLM Base, Adaptor, and Geometric Modeling. A large-scale PLM Base is employed to encode the co-evolution information in the parameters, which is used as an alternative to MSAs. Then, in Geometric Modeling, following AlphaFold2, we use modified EvoFormer and Structure Module to sufficiently exchange the information between the single representations and pair representations to capture the geometric information and recover the 3D coordinates of the atoms. We adopt an Adaptor layer to extract the co-evolution information from PLM to effectively generate the sequence and pair representations required as inputs to the Geometric modeling. The whole differentiable pipeline is trained by both self-supervised pre-training with bulks of unlabeled single sequences and supervised learning with geometric labels. 

\subsection{Large-Scale PLM Base} 
Inspired by large-scale pre-trained language models, we follow previous works on pre-training a protein language model (PLM). The PLM processes the primary protein sequences (i.e., the amino acid sequences) and extracts the knowledge needed for further geometric modeling. A protein of length $L$ can be uniquely represented by a sequence of types of amino acids denoted by $\bm{x}=(x_1, x_2, ..., x_L)$. An embedding layer $E(x_l)$ maps the type id to $d_{\textit{PLM}}$-dimension embedding vectors:
\begin{align}
\centering
\nonumber\bm{x}^{(0)} = (E(x_1), E(x_2), ..., E(x_L)).
\end{align}
Notice that $\bm{x}^{(k)} \in \mathbb{R}^{L \times d_{\textit{PLM}}}$ is the representation of the amino acid sequence.

We then apply the widely used Transformer-style blocks (\cite{vaswani2017attention} to process the embedding vectors, denoted by
\begin{align}
\bm{x}^{(k+1)}= \textit{DisentangledAttentionTransformer}(\bm{x}^{(k)}).
\end{align}
Accurately predicting the contacts between the residues, especially the long-rage contacts, is critical for protein structure prediction. Considering the contact between the residues is more dependent on the relative positions rather than the absolute positions (counted from the start of the sequence), we employ \emph{DisentangledAttentionTransformer} from DeBerTa \cite{he2020deberta} to focus on the modeling of interactions between the residue representations and the relative positions. \emph{DisentangledAttentionTransformer} adopts the attention mechanism to learn the interactions between the residues as well as the interactions of the interaction-position pairs.

Besides, we take advantage of multi-head self-attention weights in $\textit{DisentangledAttentionTransformer}$ to construct the initial pair representation. The attention weights of the $k$-th block is denoted by
$\bm{z}^{(k)} \in \mathbb{R}^{L \times L \times h_{\textit{PLM}}}$, where $h_{\textit{PLM}}$ is the number of heads of self-attention.

We add an additional \emph{Adaptor} to map the output of PLM Base to the input of Geometric Modeling module. 
\begin{align}
\centering
&\bm{\tilde{x}}^{(0)}=Linear(\bm{x}^{(n_{\textit{PLM}})}), \nonumber\\
&\bm{\tilde{z}^{(0)}}=Linear([\bm{z}^{(1)}, \bm{z}^{(2)}, \cdots, \bm{z}^{(n_{\textit{PLM}})}]),
\label{eq:adaptor}
\end{align}
where $n_{\textit{PLM}}$ is the number of blocks in PLM Base, and operator $[]$ refers to concatenation. $\bm{\tilde{x}}^{(0)} \in \mathbb{R}^{L \times d_{\textit{Single}}}$ and $\bm{\tilde{z}^{(0)}} \in \mathbb{R}^{L \times L \times d_{\textit{Pair}}}$ are the initial single representations and pair representations of the Geometric Modeling module, respectively.

\subsection{Geometric Modeling}
We employ the \emph{EvoFormer} and \emph{Structure Module} proposed by AlphaFold2 \cite{jumper2021highly} to model the relations between the residues and then estimate the 3D coordinates of the atoms in the proteins. We slightly modify the original EvoFormer and Structure Module's architecture to match our settings. First, the original EvoFormer takes the MSA representation and pair representation, encoded from the searched MSAs, as input. As an alternative, we take the output of the Adaptor (including the single representations ($\bm{\tilde{x}}^{(0)}$) and pair representations ($\bm{\tilde{z}}^{(0)}$)). Second, Evoformer adopts various attention mechanisms to exchange the information within the single and pair representations to learn the spatial relationships. Note that, compared with the original version of Evoformer proposed by AlphaFold2, we remove the column-wise gated self-attention because HelixFold-Single focuses on MSA-free protein structure prediction and is no need to exchange the messages within the MSAs. \wf{We follow the other geometric components of AlphaFold2, including} the Structure Module that takes the single representation and pair representation yielded by the EvoFormer, and exploits Invariant Point Attention and other geometric transformation operators to end-to-end predict the 3D coordinates of the atoms. \wf{Also, following} AlpahFold2, we recycle the whole Geometric Modeling module to refine the predicted structures iteratively.

\subsection{Model Optimization}
For the sake of leveraging the domain knowledge from the protein database, we operate two-stage parameter optimization on HelixFold-Single.

In the first stage, the PLM is pre-trained to capture the co-evolution information. The PLM is trained with about 300 million of single sequences recorded in a protein database. To encourage PLM to observe the diverse single sequences as soon as possible, we cluster the proteins by the similarity of single sequences and sample the proteins to balance the distributions of different clusters in our training data.
We apply the self-supervised technique masked language model (MLM) to optimize the parameters of the PLM, by randomly masking 15\% of residues in the single sequences and then reconstructing those masked residues. More concretely, MLM attempts to predict $p(x_l | x_1, ..., x_{l-1}, x_{M}, x_{l+1}, ..., x_L)$ given the residue in the $l$-th position $x_l$ being masked by $x_{M}$. A crucial proposal of this work is that PLM can learn the dependency between the masked residue and the other residues, and thus represent the co-evolution information.
Previous works \cite{rao2019evaluating} have already verified that PLMs can reveal secondary structures of the proteins, but little has been discussed on the relation between PLM and co-evolution. Co-evolution is the phenomenon that two residues in contact tend to evolve at the same time to preserve the structure and thus the function of the protein.
In PLM, if a residue at another position $s$ has a profound impact (the residue at position $s$ is changed, the masked residue will also change) on the masked residue, then those two residues are likely to evolve at the same time.

In the second stage, since merely relying on PLM to predict the structure is inadequate to capture the geometric information, PLM Base and Geometric Modeling modules in HelixFold-Single are jointly optimized. We utilize 100 thousand experimentally determined protein structures. We also use additional one million estimated protein structures for training in this stage (distilled from AlphaFold2). Following AlphaFold2, we end-to-end train the network with the main losses, including Frame Aligned Point Error (FAPE) loss and other auxiliary losses. By combining the computational efficient PLM Base module (compared with MSA search) and the Geometric Modeling module, HelixFols-Single is capable of providing efficient and precise protein structure prediction.

\section{Results}
\subsection{Datasets}
We used UniRef30 (2021-03) \cite{mirdita2017uniclust} to pre-train the PLM, which clusters UniProtKB \cite{10.1093/bioinformatics/btu739} sequences at the level of 30\% pairwise sequence identity. Then, three datasets are used to train the whole network, including the proteins in RCSB PDB \cite{10.1093/nar/28.1.235,10.1093/nar/gkaa1038} released before 2020-05-14 and two self-distillation datasets constructed from Uniclust30 (version 2018-08) and AlphaFold Protein Structure Database \cite{10.1093/nar/gkab1061}.

\subsection{Overall Comparison}
\begin{figure*}[ht]
\centering
\begin{subfigure}{0.46\columnwidth}
  \centering
  \includegraphics[width=1.0\linewidth]{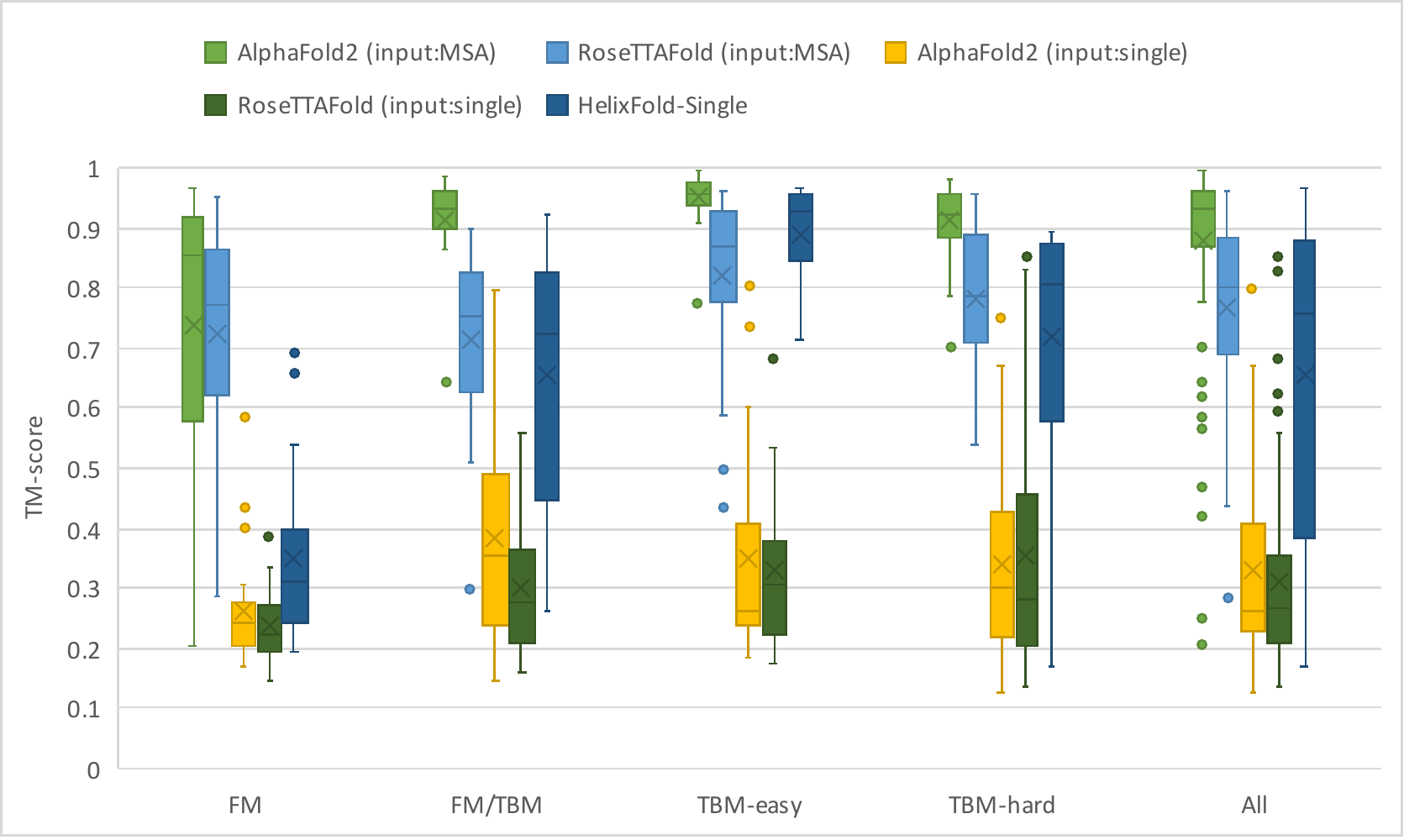}
  \caption{\emph{CASP14} (87 targets classified into FM and TBM based on their relatedness to existing structures.)}
  \label{fig:overall_casp14}
\end{subfigure}
\quad
\begin{subfigure}{0.46\columnwidth}
  \centering
  \includegraphics[width=1.0\linewidth]{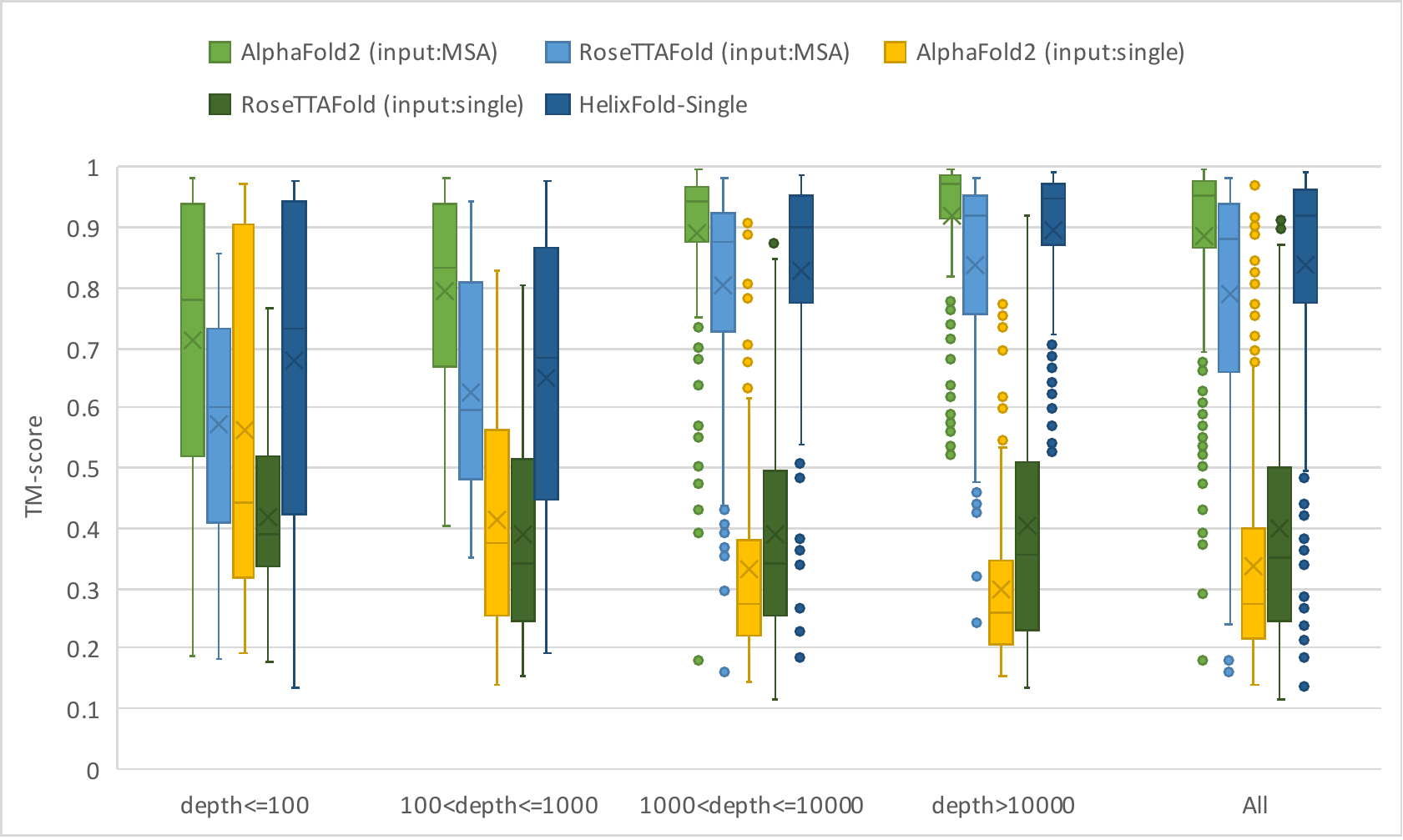}  
  \caption{\emph{CAMEO} (371 targets classified into four categories depending on different MSA depths.)}
  \label{fig:overall_cameo}
\end{subfigure}
\caption{Overall comparison of HelixFold-Single and other methods on \emph{CASP14} and \emph{CAMEO}. \emph{AlphaFold2 (input:MSA)} and \emph{RoseTTAFold (input:MSA)} are MSA-based methods, while the remaining use the primary structures as input.}
\label{fig:overall}
\end{figure*}

\emph{CASP14} \cite{jumper2021highly,https://doi.org/10.1002/prot.26202,https://doi.org/10.1002/prot.26237} with 87 domain targets and \emph{CAMEO} \cite{https://doi.org/10.1002/prot.26213} with 371 targets collected from 2021-09-04 to 2022-02-19 are used to compare the overall accuracy of HelixFold-Single with the several baseline structure prediction pipelines, including the MSA-based and MSA-free methods. AlphaFold2 \cite{jumper2021highly} and RoseTTAFold \cite{doi:10.1126/science.abj8754} are currently the most advanced methods for protein structure prediction, relying on MSAs to provide predictions. We test the accuracy of AlphaFold2 and RossTTAFold with and without homologous sequences, respectively. 
A commonly used metric, i.e., TM-score \cite{zhang2004scoring}, is exploited to evaluate the accuracy of HelixFold-Single and other methods.

Figure~\ref{fig:overall} exhibits the test results of our proposed HelixFold-Single and the compared methods on \emph{CASP14} and \emph{CAMEO}. From the results, we have the following observations:

(1) In general, HelixFold-Single significantly surpasses all the MSA-free methods on \emph{CASP14} and \emph{CAMEO} and is competitive with the MSA-based methods in some cases. Notably, the accuracy of HelixFold-Single on \emph{CAMEO} is comparable to that of \emph{AlphaFold2 (input:MSA)} and outshines another strong baseline, \emph{RoseTTAFold (input:MSA)}.  HelixFold-Single demonstrates the great potential of incorporating PLM into geometric modeling for protein structure prediction.

(2) HelixFold-Single can be par with the MSA-based methods on the targets with large homologous families, e.g., TBM-easy domain targets in \emph{CASP14} with a median of seven homologous sequences and targets with more than a thousand homologous sequences (MSA depth > 1000) in \emph{CAMEO}. These results indicate that the accuracy of HelixFold-Single is correlated to the richness of homologous sequences, revealing that the large-scale PLM adopted by HelixFold-Single is capable of embedding the information, e.g., co-evolution knowledge, of MSAs used by the MSA-based methods. 

(3) Compared HelidFold-Single with other MSA-free methods, HelixFold-Single exhibits its great superiority on all the categories of \emph{CASP14} and \emph{CAMEO}. Since AlphaFold2 and RoseTTAFold rely on MSAs as input during the training process, it is challenging for those methods to provide accurate predicts when taking only the single sequences as input.

\subsection{Effect of Number of Homologous Sequences}
\begin{figure*}[ht]
\centering
\begin{subfigure}{0.63\columnwidth}
  \centering
  \includegraphics[width=1.0\linewidth]{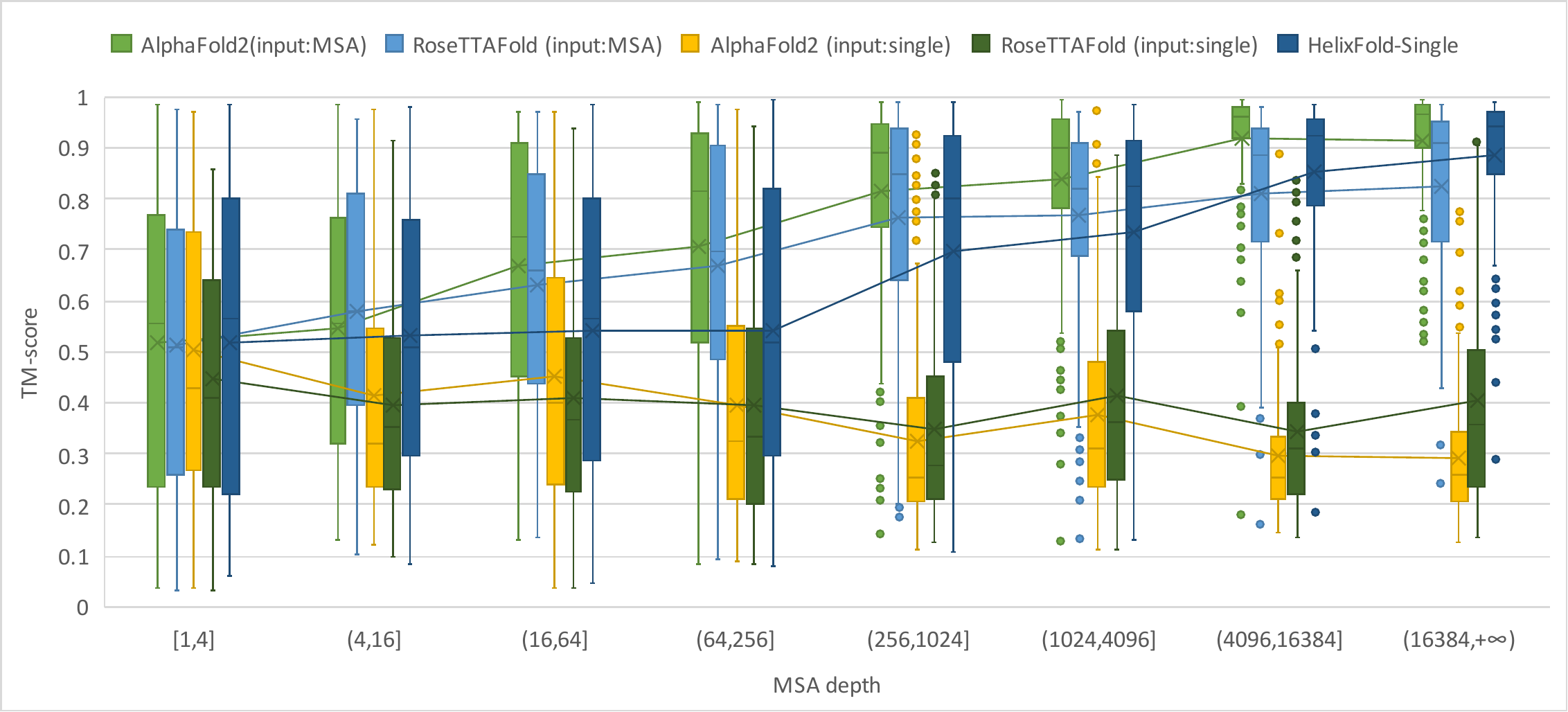}
  \caption{Comparison between HelixFold-Single and the baseline methods on protein targets with a various numbers of homologous sequences (MSA depths).}
  \label{fig:msa_depth_tmscore}
\end{subfigure}
\quad
\begin{subfigure}{0.315\columnwidth}
  \centering
  \includegraphics[width=1.0\linewidth]{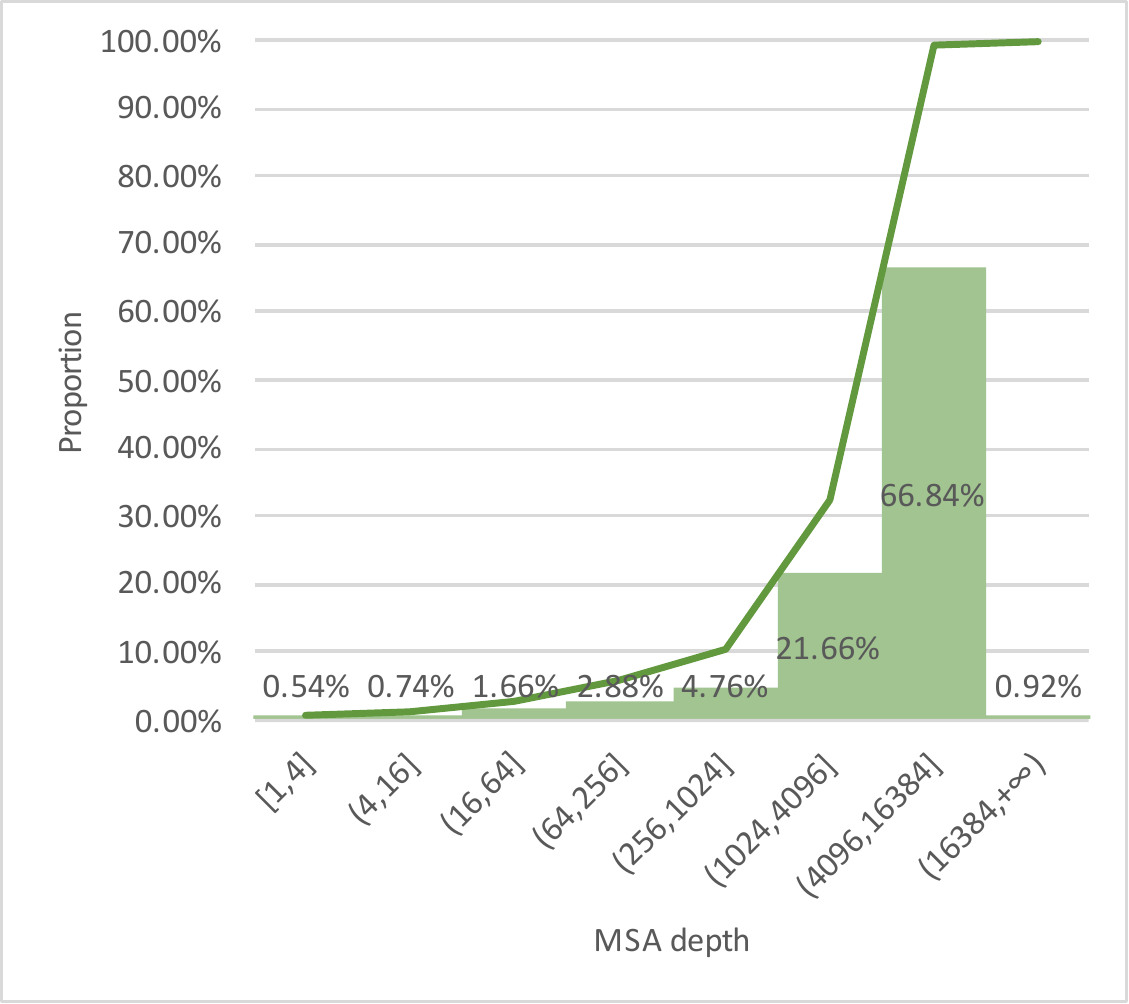}
  \caption{Distribution of proteins with different homologous sequences in PDB.}
  \label{fig:msa_depth_fre}
\end{subfigure}
\begin{subfigure}{0.32\columnwidth}
  \centering
    \includegraphics[width=1.0\linewidth]{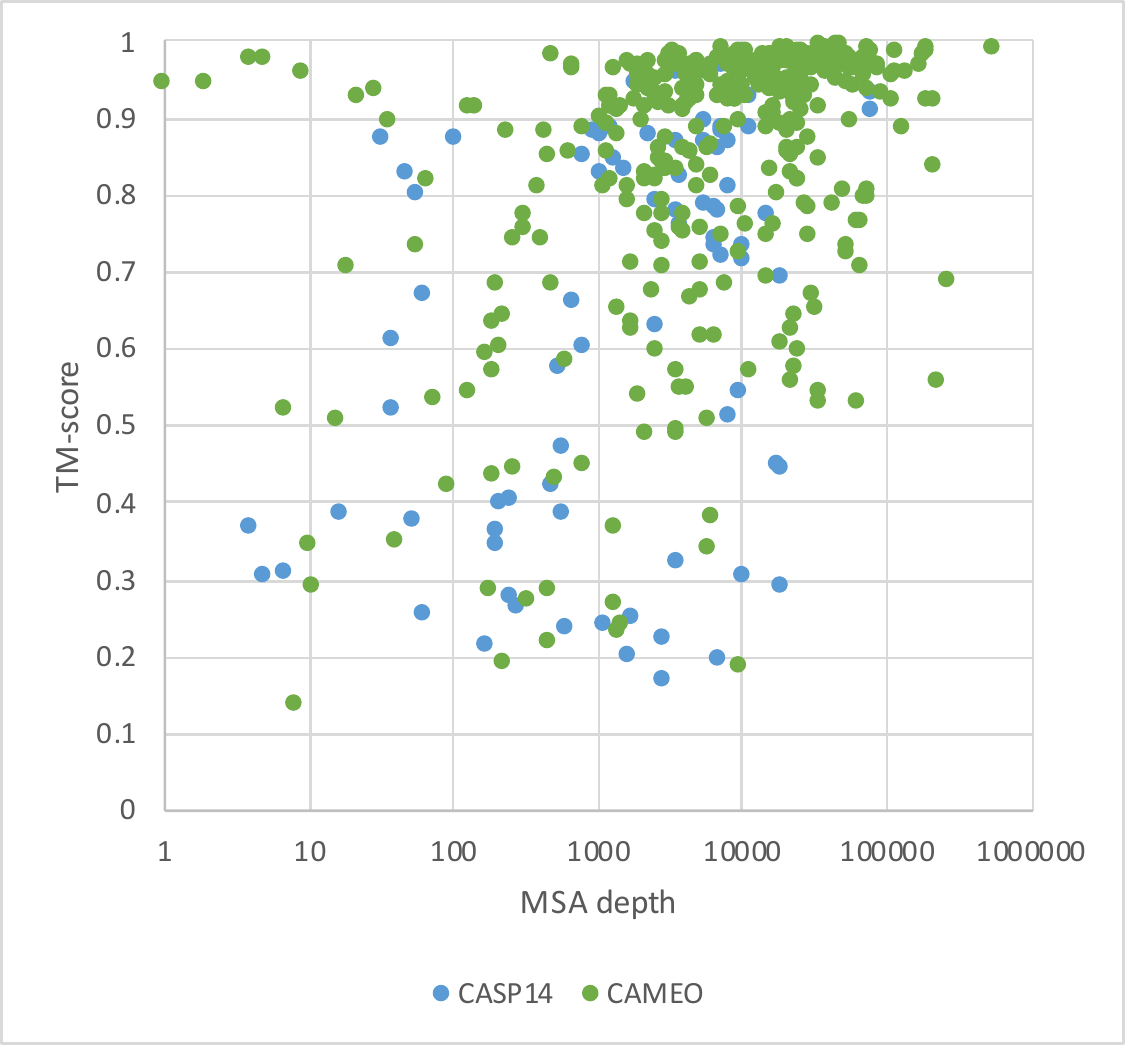}
  \caption{Relations between proteins' MSA depths and TM-scores}
  \label{fig:msa_depth_tmscore_plot}
\end{subfigure}
\begin{subfigure}{0.32\columnwidth}
  \centering
  \includegraphics[width=1.0\linewidth]{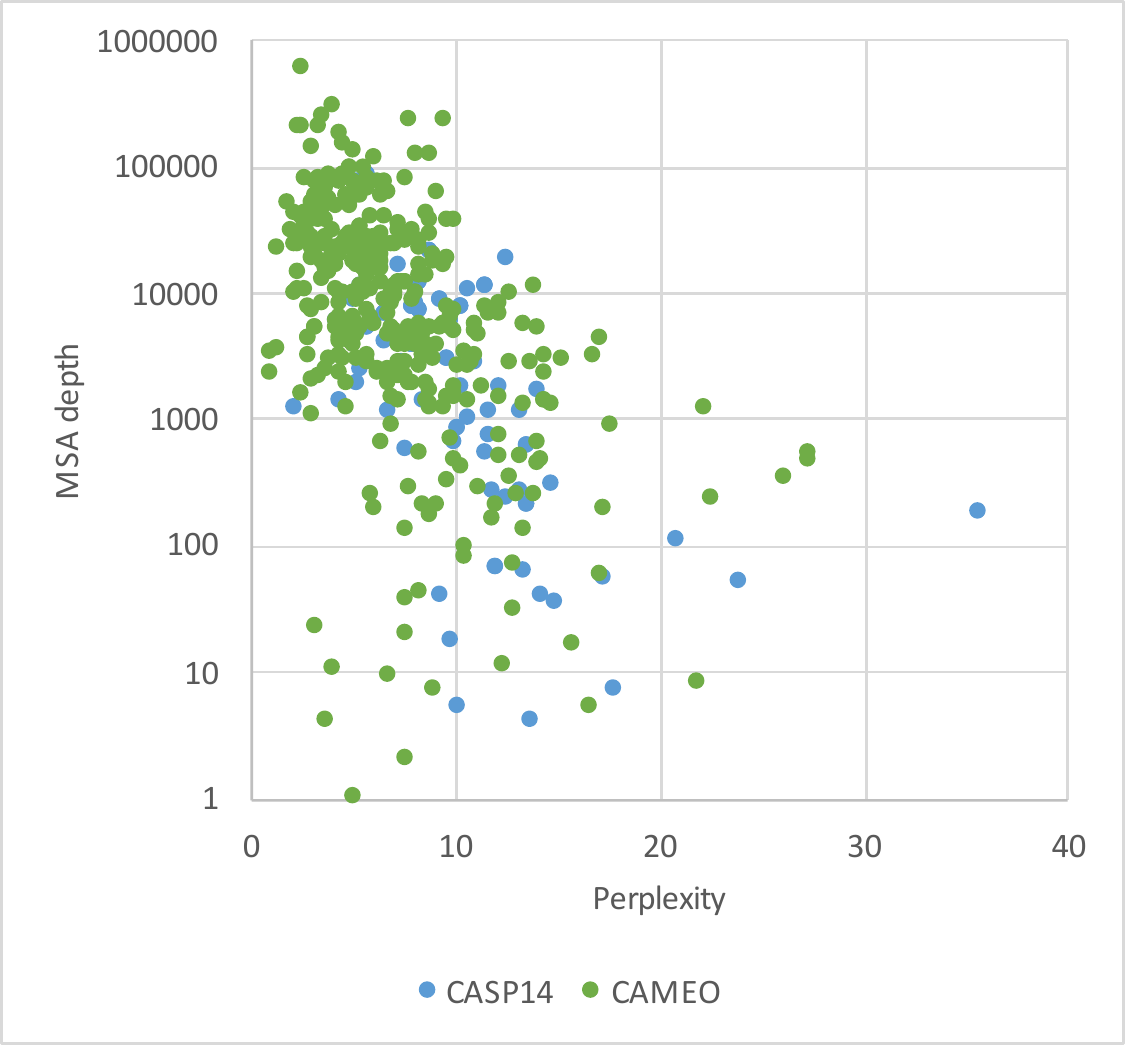}
  \caption{Relations between Perplexity of PLM and MSA depths}
  \label{fig:perplexity_msa_depth}
\end{subfigure}
\begin{subfigure}{0.32\columnwidth}
  \centering
    \includegraphics[width=1.0\linewidth]{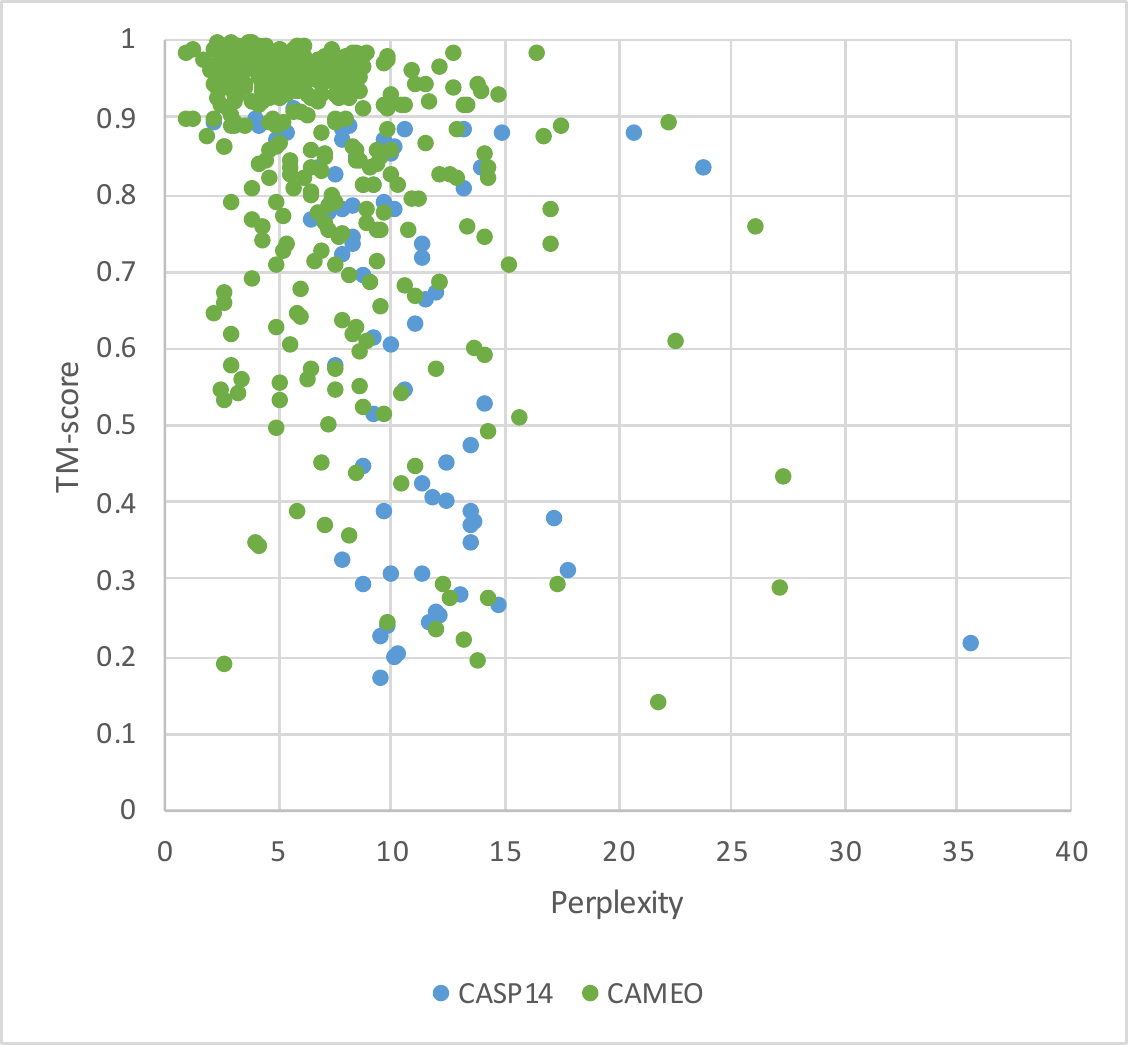}
  \caption{Relation between Perplexity of PLM and TM-scores of HelixFold-Single}
  \label{fig:perplexity_tmscore}
\end{subfigure}
\caption{Analysis of the impact of homologous sequences (MSA depths) and investigation of the relations between MSA depths, TM-scores, and perplexity of the PLM.}
\label{fig:msa_depth}
\end{figure*}

The results on \emph{CASP14} and \emph{CAMEO} indicate that the accuracy of HelixFold-Single is related to the number of homologous sequences. We further compare the performance of HelixFold-Single and other methods on the targets with variant MSA depths. We collected the targets released between 2020-05 and 2021-10 from PDB, from which we picked the targets with relatively sparse homologous sequences. We blended those targets with the data of \emph{CASP14} and \emph{CAMEO} as a new evaluation set. Figure~\ref{fig:msa_depth_tmscore} compares the TM-scores of HelixFold-Single and the baseline methods on the evaluation set, grouped by the number of homologous sequences (MSA depths). Figure~\ref{fig:msa_depth_fre} shows the distribution of the proteins in different groups in this evaluation set. We can see that as the available homologous sequences grow, the average TM-score of both HelixFold-Single and the MSA-based methods increases, while the scores of the other MSA-free methods decrease. For the proteins with sparse homologous sequences, the TM-scores of all the compared methods are unsatisfactory. For the proteins with larger homologous families, especially those with more than thousands, HelixFold-Single can compete with the MSA-based methods. Given that 90\% of the targets in PDB have more than 1024 homologous sequences, we can reasonably extrapolate that HelixFold-Single can achieve satisfying accuracy on the most frequently investigated proteins.

In order to further investigate the relationship between the capacity of the PLM, the accuracy of protein structure prediction, and the size of the homologous family, we utilized the targets in \emph{CASP14} and \emph{CAMEO} datasets to exhibit their relations, as shown in Figure~\ref{fig:msa_depth_tmscore_plot}, Figure~\ref{fig:perplexity_msa_depth}, and Figure~\ref{fig:perplexity_tmscore}. As we expected, from Figure~\ref{fig:msa_depth_tmscore_plot}, a protein's structure accuracy (TM-score) is correlated to the size of its homologous family (MSA depth), and the results are consistent with those in Figure~\ref{fig:msa_depth_fre}. Besides, we use a probability metric, Perplexity \cite{brown1992estimate}, to indicate the capacity of the protein language model. If the PLM can predict or reconstruct a protein sequence well, the Perplexity is low in predicting that target. From Figure~\ref{fig:perplexity_msa_depth} and Figure~\ref{fig:perplexity_tmscore}, we can observe that the Perplexity of the PLM and the MSA depths are negatively correlated. The Perplexity of the PLM and the TM-scores of HelixFold-Single are also negatively correlated. The results indicate that if the PLM Base module can well predict (model) a protein sequence, there is a high probability that the PLM module can learn the co-evolution information of this protein and serves as an alternative to MSAs. Thus, the Geometric Modeling module can leverage the co-evolution embedded in the PLM to provide a more accurate structure for that protein.

\subsection{Effect of the Sizes of the PLMs}
\begin{figure*}[t]
\centering
\begin{subfigure}{0.48\columnwidth}
  \centering
  \includegraphics[width=1.0\linewidth]{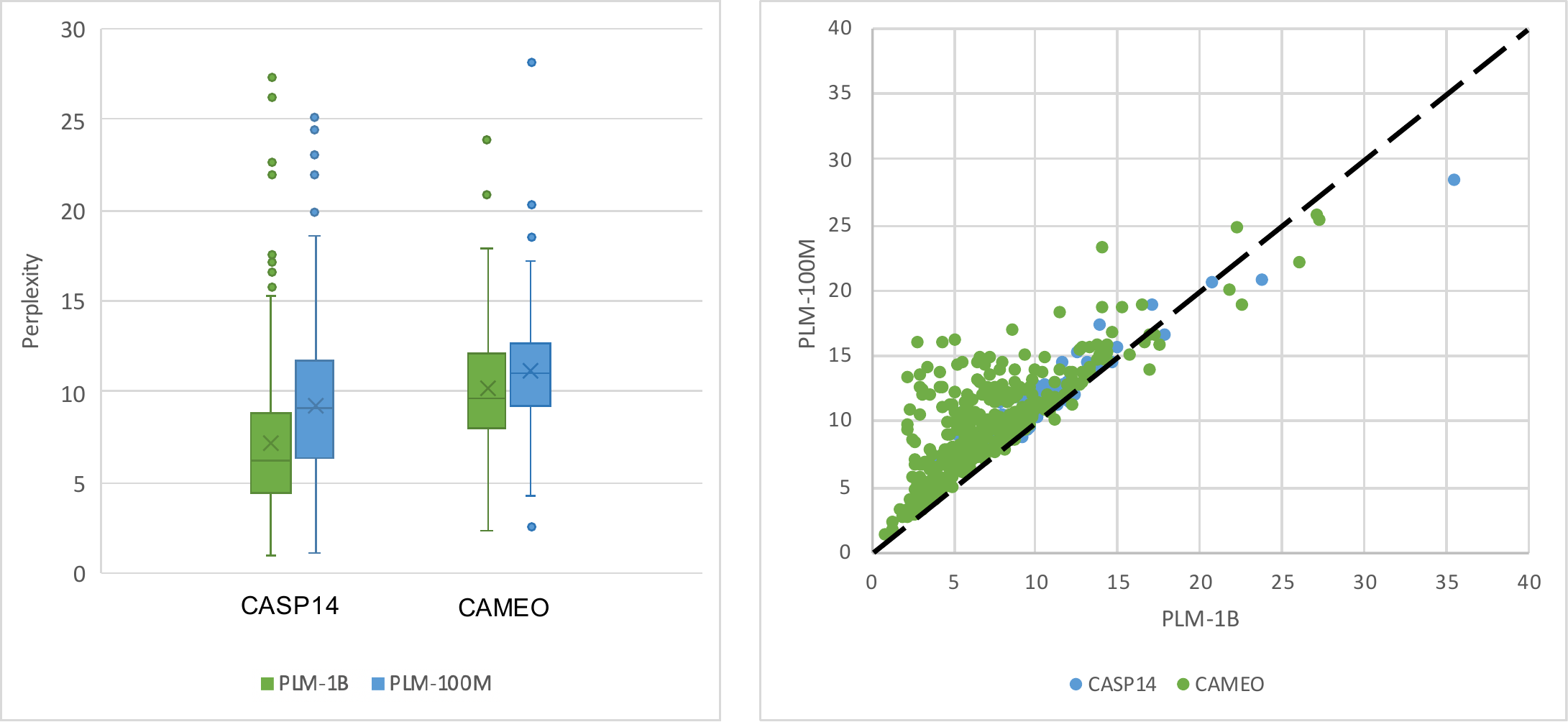}  
  \caption{Perplexity of PLM-1B and PLM-100M}
  \label{fig:PLM_perplexity}
\end{subfigure}
\quad
\begin{subfigure}{0.48\columnwidth}
  \centering
  \includegraphics[width=1.0\linewidth]{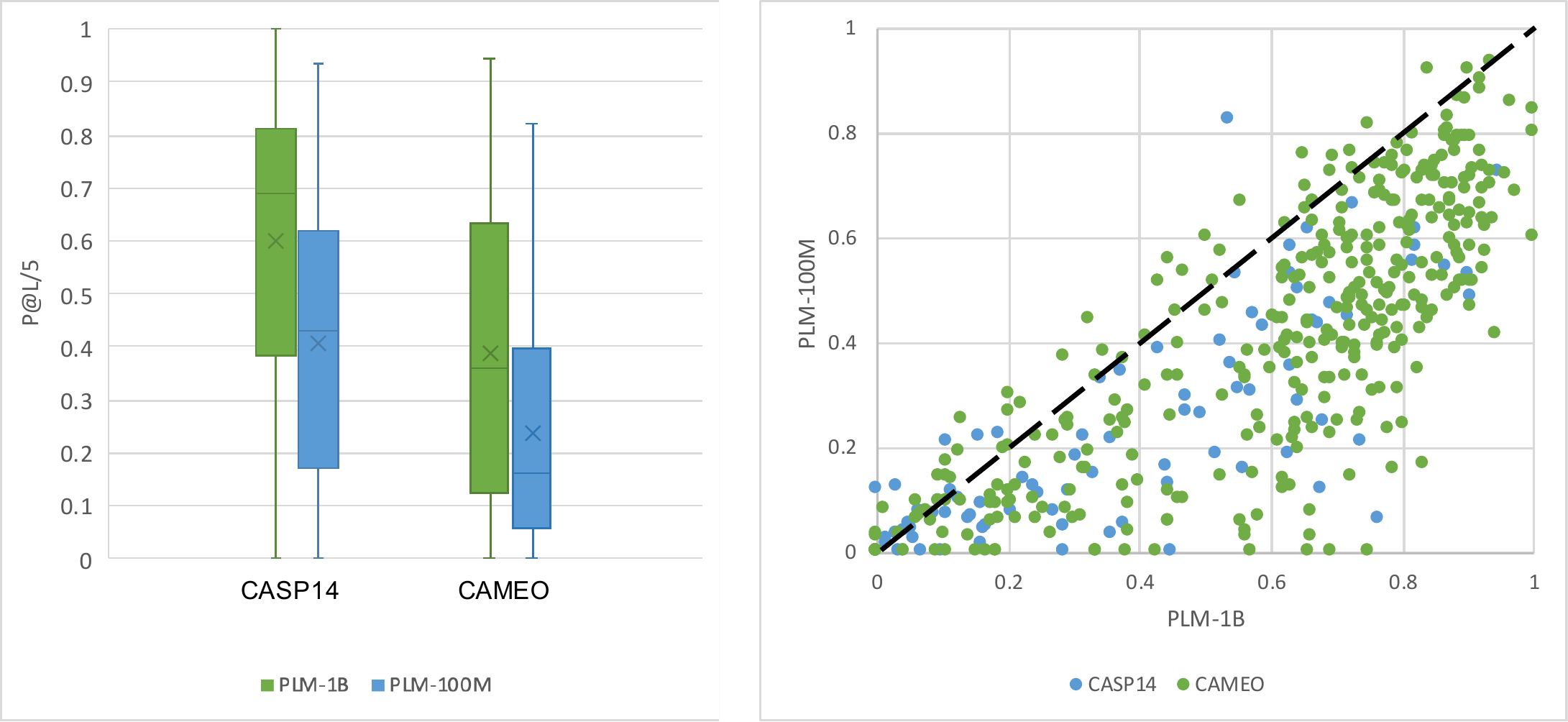}
  \caption{Contact prediction of PLM-1B and PLM-100M}
  \label{fig:PLM_contact_map}
\end{subfigure}
\caption{Comparison of PLMs of different sizes.}
\label{fig:PLM}
\end{figure*}
To comprehensively study the ability of the PLMs of different sizes to learn the co-evolution information, we compare a pre-trained PLM of 1B parameters (denoted by PLM-1B) and another pre-trained PLM of 100M (denoted by PLM-100M). Table~\ref{fig:PLM_perplexity} exhibits the Perplexity of PLM-1B and PLM-100M of the targets from datasets \emph{CASP14} and \emph{CAMEO}. In general, the smaller the perplexity is, the stronger the capacity of the PLM is. Thus, PLM-1B with more model parameters performs better than PLM-100M with fewer parameters on both datasets \emph{CASP14} and \emph{CAMEO}. In addition, we apply the PLM-1B and PLM-100M on the task of protein residue contact prediction to compare their performance on the downstream tasks. We simply fit a logistic regression that takes the attention weights, i.e., $[\bm{z}^{(1)}, \bm{z}^{(2)}, \cdots, \bm{z}^{(n_{\textit{PLM}})}]$, from the PLMs as input and predict the contact of residues on the targets in datasets \emph{CASP14} and \emph{CAMEO}. Following \cite{rao2019evaluating,rao2021msa}, we use top L/5 long-range contact precision, denoted by P@L/5, as the evaluation metric, and the results are shown in Figure~\ref{fig:PLM_contact_map}. As we can see, PLM-1B is significantly superior to PLM-100M on the contact prediction task. The results from Figure~\ref{fig:PLM_perplexity} and Figure~\ref{fig:PLM_contact_map} both support the hypothesis that the larger the size of the PLM is, the stronger its capacity is. Therefore, it can be reasonably inferred that the performance of the PLM will continue to improve as the size of the PLM increases to a larger size.

\subsection{Prediction Speed Comparison}
\begin{figure}[ht]
\centering
\includegraphics[width=0.5\linewidth]{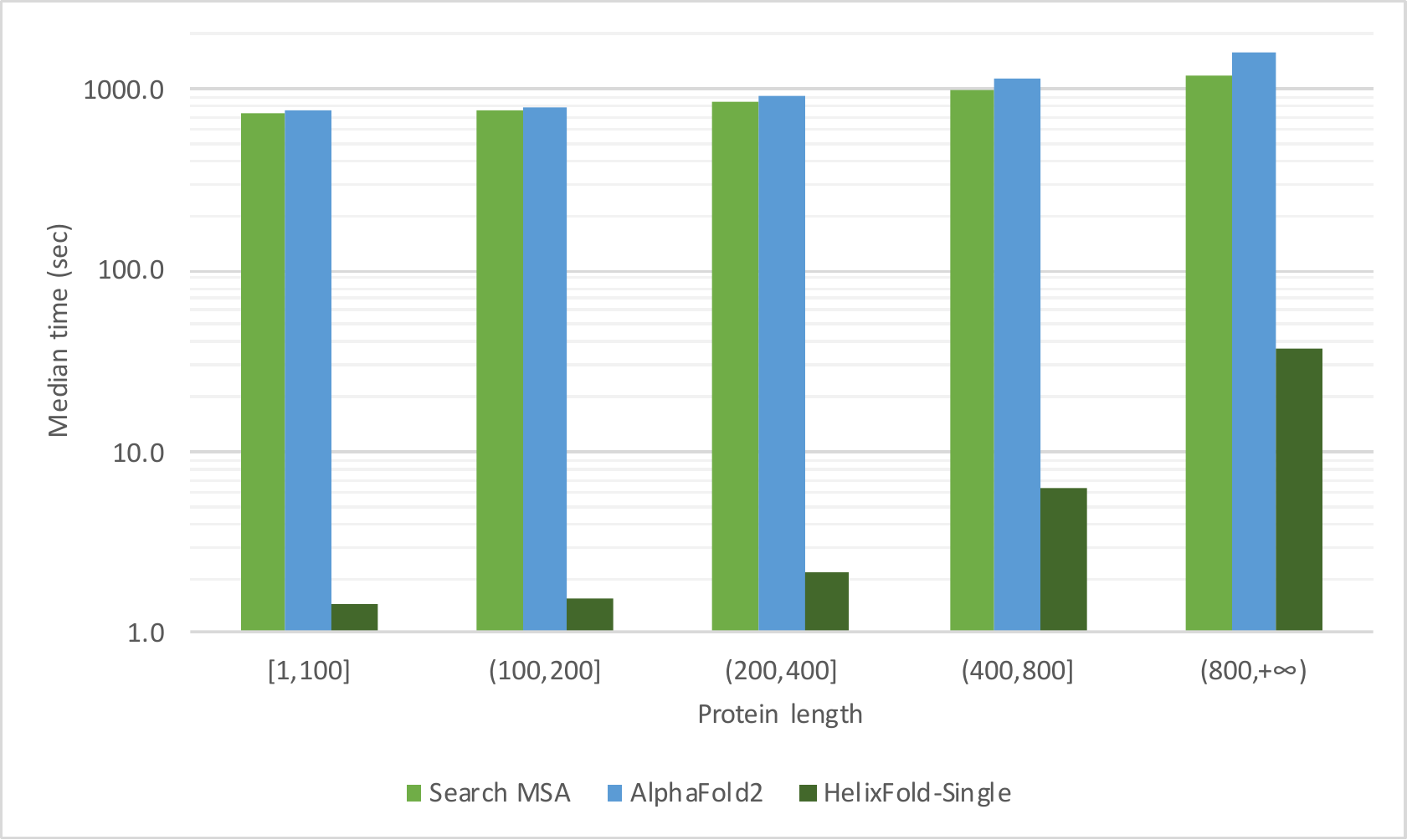}  
\caption{Median times of MSA search, AlphaFold2, and HelixFold-Single on proteins with various lengths.}
\label{fig:efficiency}
\end{figure}
Massive time consumption for searching MSAs is one of the bottlenecks of the MSA-based folding, and accelerating the speed of protein structure prediction can considerably broader its applications. The MSA-free HelixFold-Single has a tremendous advantage for inference efficiency for exempting MSA searching. Figure~\ref{fig:efficiency} exhibits the computation time cost of 1. MSA searching; 2. Whole inference pipeline of AlphaFold2; 3. Inference of HelixFold-Single. All the tests are executed in a single NVIDIA A100(40G) GPU. In general, Helixfold-Single consumes much less time than the Alphafold2, while AlphaFold2 pipeline spends most of its time in MSA searching. For proteins less than 100 in length, HelixFold-Single's prediction time is only about one-thousandth of that of AlphaFold2. Even for the proteins with more than 800 amino acids, HelixFold-Single still has great efficiency superiority. The high efficiency of HelixFold-Single demonstrates the potential of its application in tasks with a great demand for structural prediction.

\subsection{Case Study}
\begin{figure*}[ht]
\centering
\begin{subfigure}{0.23\columnwidth}
  \centering
  \includegraphics[width=1\linewidth]{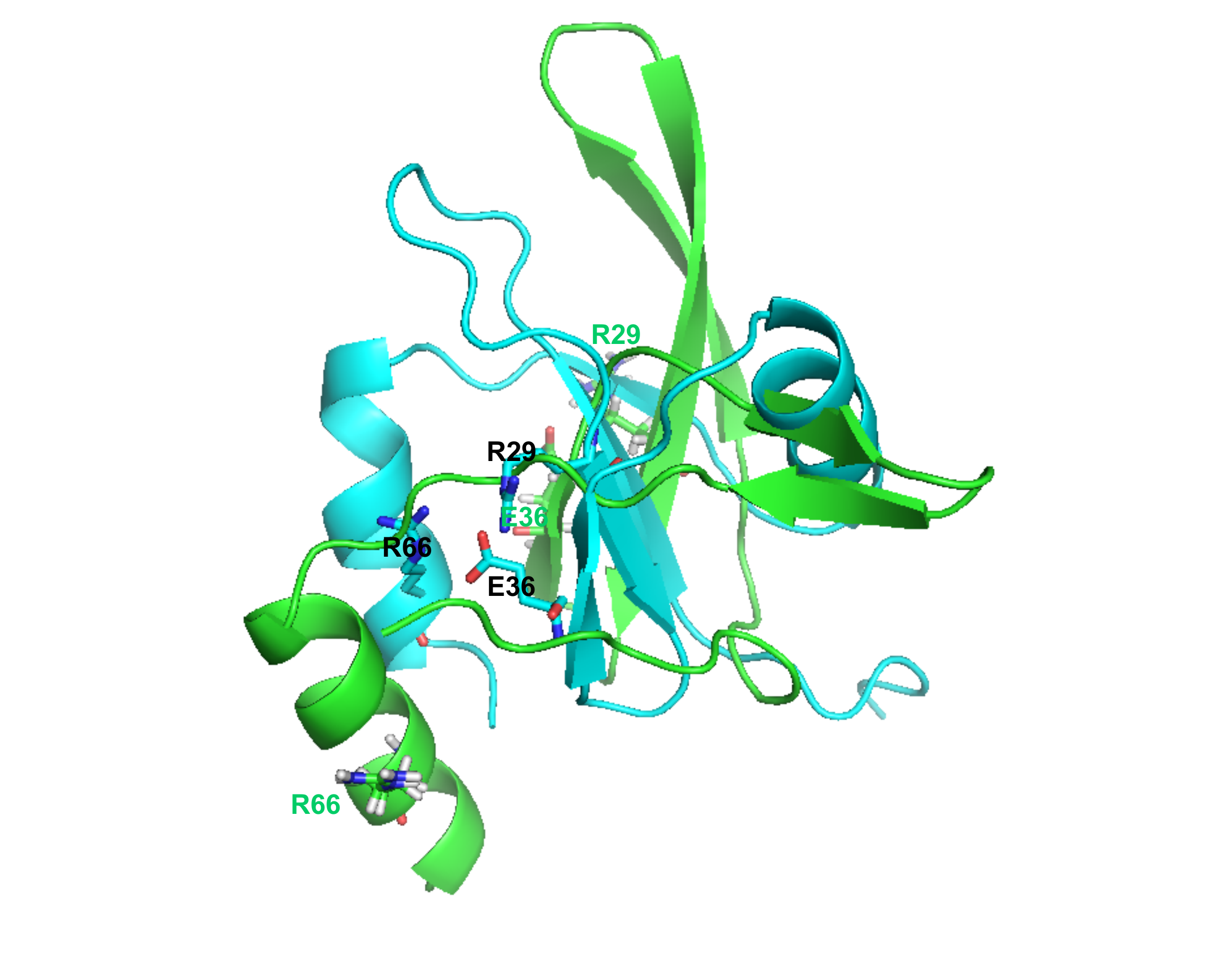}  
  \caption{7KWT:B, AlphaFold2, TM-score=0.2425}
  \label{fig:7kwt_b_af2_exp}
\end{subfigure}
\quad
\begin{subfigure}{0.23\columnwidth}
  \centering
  \includegraphics[width=1\linewidth]{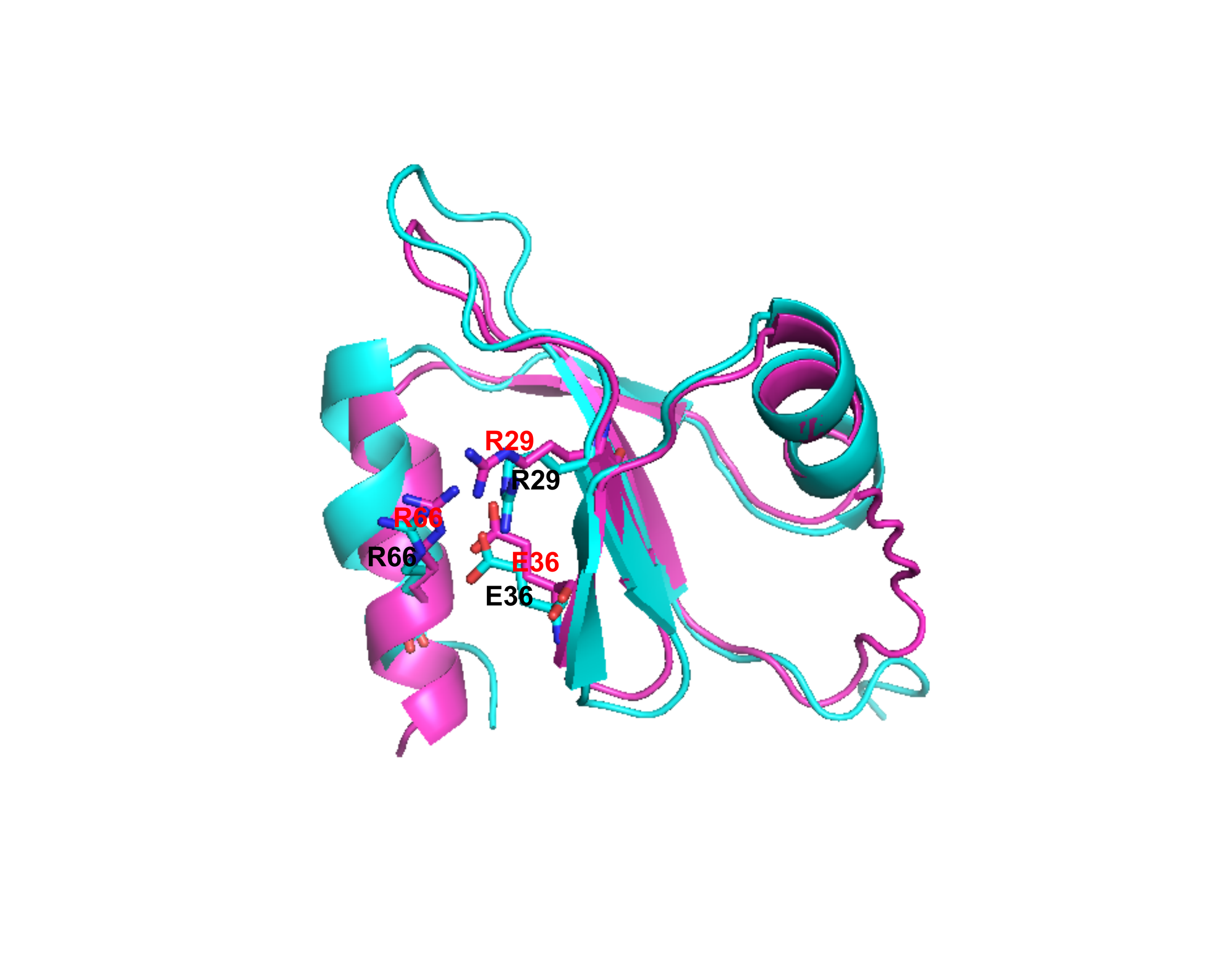}  
  \caption{7KWT:B, HelixFold-Single, TM-score=0.8066}
  \label{fig:7kwt_b_helix_exp}
\end{subfigure}
\quad 
\begin{subfigure}{0.23\columnwidth}
  \centering
  \includegraphics[width=1\linewidth]{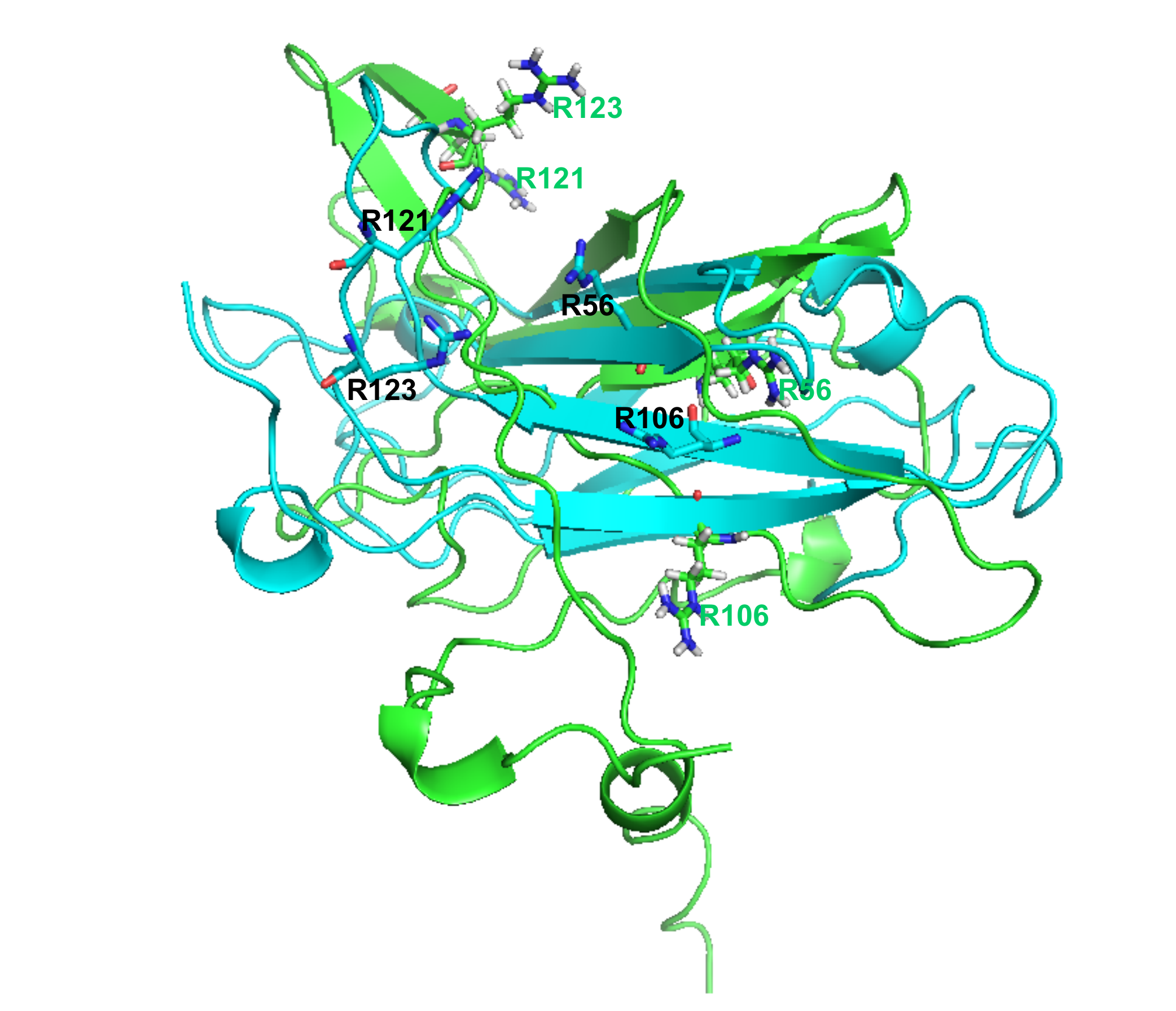}  
  \caption{7BCJ:A, AlphaFold2, TM-score=0.2914}
  \label{fig:7kww_a_af2_exp}
\end{subfigure}
\quad
\begin{subfigure}{0.23\columnwidth}
  \centering
  \includegraphics[width=1\linewidth]{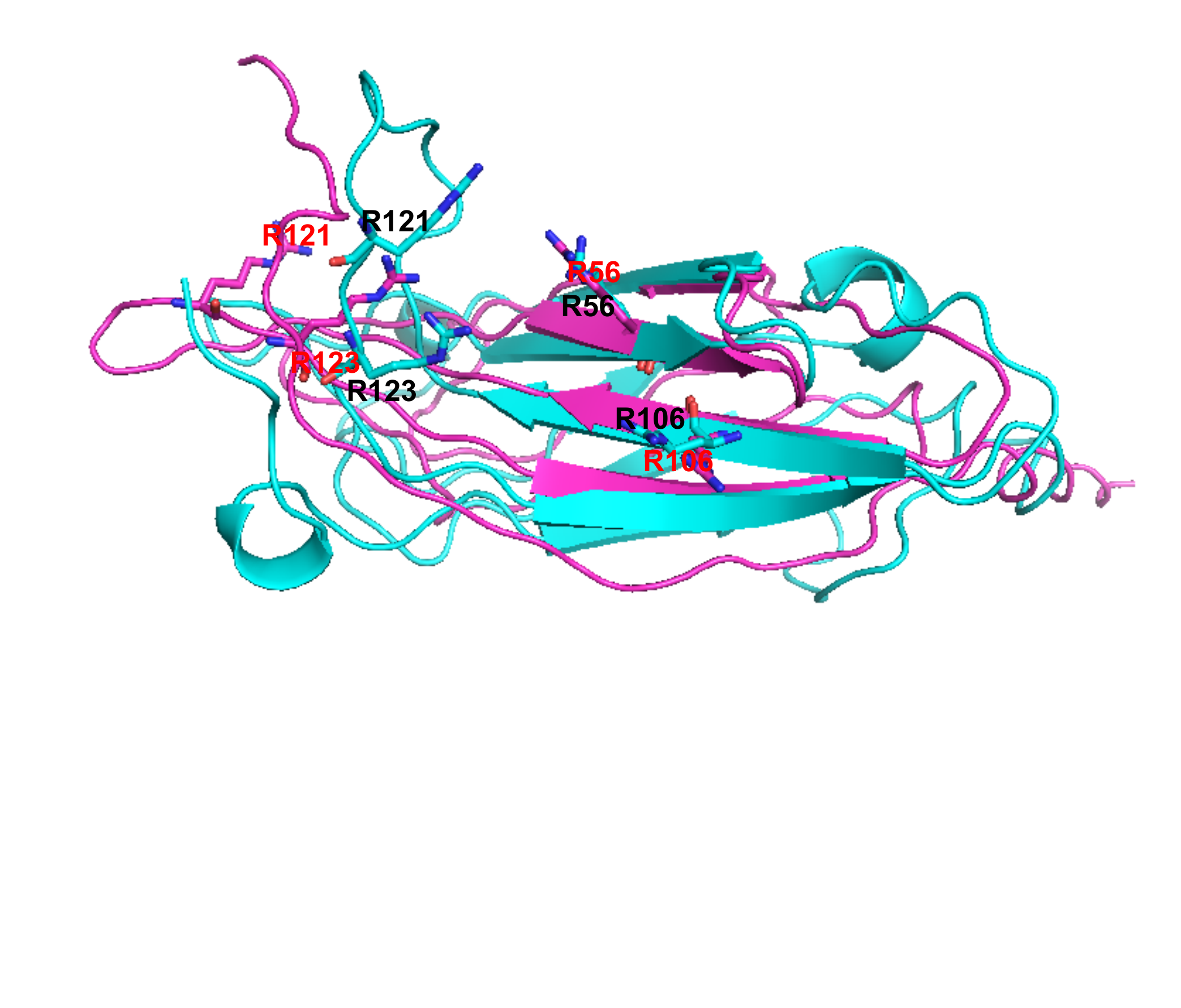}  
  \caption{7BCJ:A, HelixFold-Single, TM-score=0.6420}
  \label{fig:7kww_a_helix_exp}
\end{subfigure}
\caption{HelixFold-Single predicts PlyC and RoxP structure more accurately than AlphaFold2. PlyC structures predicted by (a) AlphaFold2 and (b) HelixFold-Single is aligned with the reference structure (PDB ID: 7KWT, chain B); RoxP structure predicted by (c) AlphaFold2 and (d) HelixFold-Single is aligned with the reference structure (PDB ID: 7BCJ, chain A). A-D) Green: structure predicted by AlphaFold2. Magentas: structure predicted by HelixFold-Single. Cyan: reference crystal structure measured by X-RAY diffraction approach (resolution<1.8A). Key residues related to protein function are shown as sticks.}
\label{fig:case_study}
\end{figure*}
Most proteins exert their functions by interacting with other molecules. Changes in the structure of a protein, especially those in the key interacting residues, can significantly affect its biological function. As a result, a protein’s function is closely associated with its structure, and accurately predicting the structure would facilitate our understanding of its biological role. While AlphaFold2 achieves outstanding accuracy in most of the protein structure prediction tasks, its performance can still be poor in some situations. Here, we demonstrate that HelixFold-Single complements AlphaFold2 in several of these cases. Endolysin enzymes from bacteriophages cause bacterial lysis by degrading the peptidoglycan cell wall. The streptococcal C1 phage endolysin PlyC is the most potent endolysin and can rapidly lyse group A, C, and E streptococci. Study on PlyC structure revealed that the key residues, including R66, E36, R29, etc, are important for the binding of PlyC to its target and hence are critical to its function [26]. However, AlphaFold2 failed to produce the reliable structure of the protein (Figure~\ref{fig:case_study}(a)). This is probably due to insufficient co-evolution information extracted from MSAs. In contrast, the structure predicted by HelixFold-Single (Figure~\ref{fig:case_study}(b)) more closely resembles the one measured by the experiment, likely attributed to its little dependence on the information from MSAs. A similar result is observed for another protein RoxP. This protein is produced by Cutibacterium acnes, a predominant bacterium on human skin, and was shown to alleviate radical-induced cell damage. The key residues R56, R106, R121, R123 on RoxP form a positively charged groove, which acts as the binding site for substrate and cofactors [27]. HelixFold-Single accurately predicts the formation of the positively charged groove(Figure~\ref{fig:case_study}(d)), which is not observed in the structure predicted by AlphaFold2 (Figure~\ref{fig:case_study}(c)). Furthermore, the TM-score of HelixFold-Single for RoxP is much higher than that of AlphaFold2, suggesting an overall better performance of HelixFold-Single in predicting RoxP structure. Altogether, our case studies indicate that HelixFold-Single outperforms AlphaFold2 in some situations and can be used as a reliable tool to analyze the function of proteins without known X-RAY structures.

\section{Related Works}
\subsection{Protein Language Models}
Large-scale language models \cite{vaswani2017attention} with the self-supervised learning (SSL) paradigm, such as masked language model (MLM) \cite{kenton2019bert} and auto-regression \cite{radford2018improving}, have achieved extraordinary success in Natural Language Processing (NLP) tasks. Recent progress has revealed that their capabilities are deeply related to the scale of the model parameters: the larger the scale of the parameters, the better the performance \cite{brown2020language}. The community has not yet seen a sign of stopping growth by moving from billions to hundreds of billions of parameters. Those language models are capable of memorizing and generalizing massive common-sense knowledge and professional expertise implicitly included in the large-scale unlabeled data. Inspired by those achievements, Protein Language Models (PLMs) tried to transfer language models and SSL tasks to protein modeling. A protein can be represented by an amino acid sequence, similar to the sequences of words or tokens in NLP. Previous works \cite{rao2019evaluating, elnaggar2020prottrans, rao2020transformer, xiao2021modeling} have shown that by pre-training with only single sequences without much supervision, protein language models can reveal the protein classification, stability, and lower-level structure information (including secondary, tertiary structures and 2D contact maps). However, the accuracy of these models in structure prediction is still far from that of the mainstream folding models supervised by the ground-truth protein structure.

\subsection{Protein Structure Prediction}
Mainstream pipelines \cite{yang2020improved,yang2015tasser,du2021trrosetta,peng2011raptorx} rely on extracting the co-evolution information from Multiple Sequence Alignments (MSAs) to predict the protein structures. Earlier works manually designed the features derived from MSAs, such as inverse covariance matrices of MSAs. Then, deep neural networks (DNNs), e.g., convolutional networks, are utilized to model the relations between the residues. Advanced studies \cite{jumper2021highly,du2021trrosetta}, directly take the MSAs as input and apply DNNs to predict the 3D coordinates of the proteins. Particularly, the appearance of AlphaFold2 \cite{jumper2021highly} has dramatically narrowed the accuracy gap between the experimentally determined structures and model estimated structures, employing the EvoFormer module to enhance the interaction between MSA sequences and pairwise geometric information and the Structure module to directly predict the atoms' coordinates. However, the reliance on MSA inevitably impedes the computation efficiency and accurate prediction of orphan proteins and designed proteins, as well as downstream tasks such as protein design. 

Although the structure of a protein is dependent on its primary structure, it is incredibly challenging to train an accurate model that can infer the protein structures with only the primary structures. Only a small number of samples, i.e., experimentally determined structures recorded in the PDB database, are available for model training. Several works attempt to incorporate the protein language models (PLMs) for MSA-free protein structure prediction. RGN2 \cite{chowdhury2021single} employs a protein language model (AminoBERT) with a recurrent geometric network that utilizes Frenet-Serret frames to generate the backbone structure. Besides, advanced studies \cite{weissenow2022protein, wang2022single} combine pre-trained PLMs, such as ProT5 \cite{elnaggar2020prottrans} and ESM-1b \cite{rives2021biological}, with ResNets to predict 2D structures, e.g., contact map of a protein, yielding superior performance in orphan proteins. Nonetheless, the overall accuracy of those works is still unsatisfactory due to the limited capacity of the used model architectures.

\section{Conclusion and Future Work}
On the one hand, mainstream protein structure prediction methods, such as AlphaFold2 and RoseTTAFold, rely on the MSAs to extract the homologous information. However, searching MSAs is time-consuming, limiting the application of those methods to broader protein-related tasks. On the other hand, the large-scale protein language model learns the protein correlations from a great number of unlabeled proteins through self-supervised learning tasks. By utilizing large-scale parameters to embed the homologous information, we prove it can be used as an alternative to MSAs to reduce the time consumption required by the protein structure prediction methods. HelixFold-Single attempts to take advantage of both the protein language model and the geometric modeling, end-to-end predicting the protein structures with only the primary structures. HelixFold-Single can be par with the MSA-based methods on targets with large homologous families and is much more efficient than the MSA-based methods, demonstrating its application prospect for protein study.

In the future, as the experimental results indicate that the larger size of the PLM can achieve superior performance, we will continue investigating the PLM with a larger size for protein structure prediction. In addition, the accuracy of the targets with only a few homologous sequences is still unsatisfactory. Thus we will try to introduce more diverse training data to alleviate this problem.

\section{Data Availability}
To pre-train the PLM, UniRef30 (2021-03) is publicly available in \url{https://wwwuser.gwdg.de/~compbiol/uniclust/2021_03/}. While to train the whole network, RCSB PDB can be downloaded in \url{https://www.rcsb.org/docs/programmatic-access/file-download-services} and AlphaFold Protein Structure Database as the distillation dataset can be downloaded in \url{https://ftp.ebi.ac.uk/pub/databases/alphafold/v2/}.

\section{Code Availability}
The source code, trained weights and inference code of HelixFold-Single are freely available at GitHub \url{https://github.com/PaddlePaddle/PaddleHelix/tree/dev/apps/protein_folding/helixfold-single} to ensure the reproduction of our experimental results.

\clearpage

\section*{Appendix A: Training and Evaluation Data}
\subsection*{Training Data}
UniRef30 (2021-03) \cite{mirdita2017uniclust}, containing about 260 millions protein sequences is utilized to pre-train the PLM, clustering UniProtKB \cite{10.1093/bioinformatics/btu739} sequences at the level of 30\% pairwise sequence identity.

Three datasets are utilized to train HelixFold-Single for MSA-free protein strcture prediction.
\begin{itemize}
    \item \textbf{RCSB PDB \cite{10.1093/nar/28.1.235,10.1093/nar/gkaa1038}:} The targets released before 2020-05-14 in PDB are used to train HelixFold-Single. We filter out the targets with resolution larger than 3Å and whose number of amino acids less than 10. The targets are clustered at 40\% sequence identity cutoff.
    \item \textbf{Distillation-Uniclust30:} We inference the structures of the targets in Uniclust30 (version 2018-08) by AlphaFold2 for self-distillation. We follow the data-prepossess procedure reported in AlphaFold2. Further, the target structures with average pLDDT less than 0.5 are filtered out. Then, the targets are clustered at 30\% sequence identity cutoff.
    \item \textbf{Distillation-EBI:} About one million protein structures are extracted from AlphaFold Protein Structure Database \cite{10.1093/nar/gkab1061}. We removed the protein structures with average pLDDT less than 0.5. The remaining targets are clustered at 50\% sequence identity cutoff.
\end{itemize}

\subsection*{Evaluation Data}
We exploit three datasets to evaluate the accuracy and efficiency of HelixFold-Single and the baseline methods.
\begin{itemize}
    \item \textbf{CASP14:} 61 targets are collected from CASP14 \cite{https://doi.org/10.1002/prot.26202,https://doi.org/10.1002/prot.26237} for overall evaluation, which includes 87 domains with classification of FM (free modeling), TBM-easy (easy template-based modeling), TBM-hard (hard template-based modeling) and FM/TBM (modeling with only remote structural homologies).
    \item \textbf{CAMEO:} We collect 371 targets from CAMEO \cite{https://doi.org/10.1002/prot.26213} between 2021-09-04 and 2022-02-19, which consists of various target difficulties including Easy, Medium, and Hard.
    \item \textbf{MSA Depth Test:} We create a test set obtained from RCSB PDB, including 793 targets with a wide range of different MSA depths from 2020-05 to 2021-10, especially the targets with only a few homologous sequences. This test set is combined with datasets \emph{CASP14} and \emph{CAMEO} to investigate the effect of the number of homologous sequences.
\end{itemize}

\section*{Appendix B: Detailed Settings of HelixFold-Single}
\subsection*{Training Settings}
The implementation of HelixFold-Single is based on our previous work, HelixFold \cite{wang2022helixfold}, and we use 128 NVIDIA A100 GPUs to train HelixFold-Single.
Table~\ref{tab:architecture} exhibits the architecture setting of HelixFold-Single. We train two version of PLMs for ablation study. To balance the computation costs of multiple GPUs for pre-training, the batch size used in each GPU is dynamically adjusted according to the lengths of protein sequences. We use AdamW optimizer \cite{loshchilov2018decoupled} with learning rate of 5e-4, $\beta_1$ = 0.9, $\beta_2$ = 0.999, weight decay of 0.01, learning rate warm-up over the first 30,000 steps. When training the whole network for protein structure estimation, we use Adam optimizer \cite{kingma2014adam} to optimize the parameters. We apply two stages of training: initial training stage and fine-tuning stage. In the initial training stage, the learning rate is set to be 1e-3 and the lengths of the input protein sequences are cropped to be 256. In the fine-tuning stage, we use learning rate of 2e-4 and the lengths of the input protein sequences are cropped to be 384. Gradient clipping by the global norm \cite{pascanu2013difficulty} is adopted on the parameters with a clipping value of 1.0.

\begin{table}[ht]
 \caption{Architecture setting of HelixFold-Single.}
  \centering
    \begin{tabular}{lccccc}
    \toprule
    Components & Model size & \tabincell{c}{Layer num} & \tabincell{c}{Hidden size} & Intermediate size & \tabincell{c}{Head num} \\ 
    \midrule
    PLM-1B & 1.09B & $n_{\textit{PLM}}=20$ & $d_{\textit{PLM}}=2048$ & 8192 & $h_{\textit{PLM}}=16$ \\
    PLM-100M & 100M & $n_{\textit{PLM}}=12$ & $d_{\textit{PLM}}=768$ & 3072 & $h_{\textit{PLM}}=12$ \\
    \midrule
    \midrule
    EvoFormer & 87M & $n_{\textit{EvoFormer}}=24$ &
    \tabincell{c}{$d_{\textit{Single}}=512$\\$d_{\textit{Pair}}=64$} & & \\
    \midrule
    Structure Module & 1.7M & $n_{\textit{Structure}}=8$ & $d_{\textit{Structure}}=384$ & & \\
    \bottomrule
    \end{tabular}
    \label{tab:architecture}
\end{table}

\subsection*{Model Architecture}
\subsubsection*{PLM Base}
\begin{figure}[ht]
\centering
\includegraphics[scale=0.3]{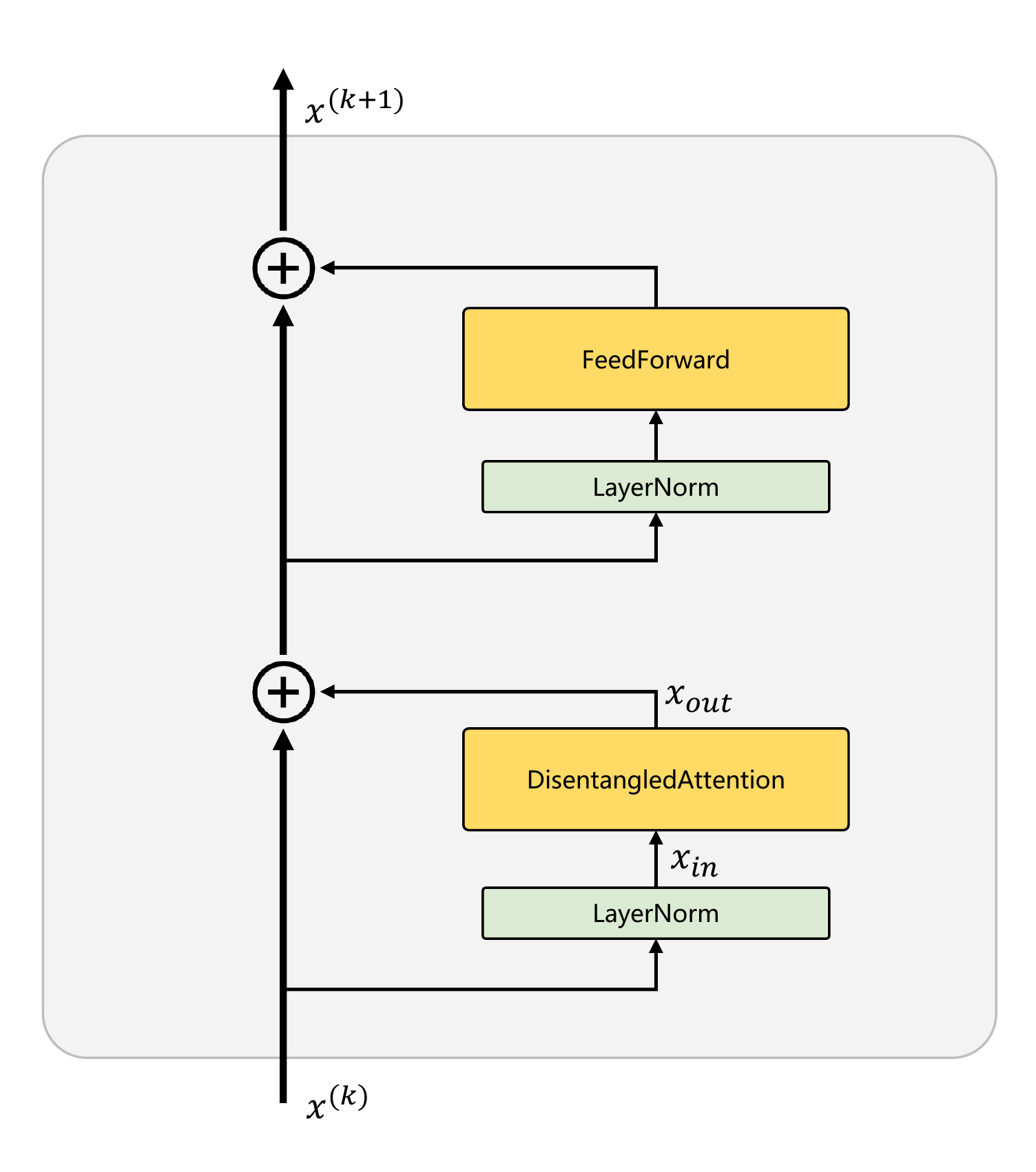}  
\caption{Architecture of \emph{DisentangledAttentionTransformer} (superscript $k$ denotes the layer id).}
\label{fig:deberta}
\end{figure}

As shown in Figure~\ref{fig:deberta}, PLM Base is mainly based on DeBerTa \cite{he2020deberta}. We make two slight modifications: (1) To stabilize the pre-training of PLM, instead of using Post-Norm in DeBerTa, Pre-Norm \cite{xiong2020layer} is applied in PLM Base of HelixFold-Single. (2) We find that using residue-to-position and residue-to-residue (Equation~\ref{eq_deberta}) is enough, while the performance gain by adding position-to-residue is trivial. Thus, we left out the position-to-residue term in DeBerTa. As a result, we have the $DisentangledAttention$ layer denoted by

\begin{equation}
\begin{aligned}
\boldsymbol{q}=\boldsymbol{x}_{in} \boldsymbol{W}_{\boldsymbol{q}}, \quad \boldsymbol{k}=\boldsymbol{x}_{in} \boldsymbol{W}_{\boldsymbol{k}}, \quad \boldsymbol{v}=\boldsymbol{x}_{in} \boldsymbol{W}_{\boldsymbol{v}}, \quad \boldsymbol{p}=\boldsymbol{e}_p \boldsymbol{W}_{\boldsymbol{p}} \\
\begin{array}{l}
A_{i, j}=\underbrace{\boldsymbol{q}_{i} \boldsymbol{k}_{j}^{\mathsf{T}} }_{\text {(a) residue-to-residue }}+\underbrace{\boldsymbol{q}_{i} \boldsymbol{p}_{\delta(i, j)}^{\mathsf{T}}}_{\text {(b) residue-to-position }}  \text{ } \text{ } \text{ } \text{ } \text{ } \text{ } \text{ } \text{ } \text{ } \text{ } \text{ } \text{ } \text{ } \text{ } \text{ } \text{ } \text{ } \text{ } \text{ }   \\
\boldsymbol{x}_{out}=\operatorname{softmax}(\frac{\boldsymbol{A}}{\sqrt{2 d_{\textit{PLM}}}}) \boldsymbol{v}
\label{eq_deberta}
\end{array}
\end{aligned}
\end{equation}

$\boldsymbol{W}_{\boldsymbol{*}}$ are trainable parameters; $\boldsymbol{e}_p$ is the trainable position embedding; $\delta(i, j)$ denotes the relative distance between position $i$ and $j$.

\subsubsection*{Geometric Modeling}
\begin{figure}[ht]
\centering
\includegraphics[width=1.0\linewidth]{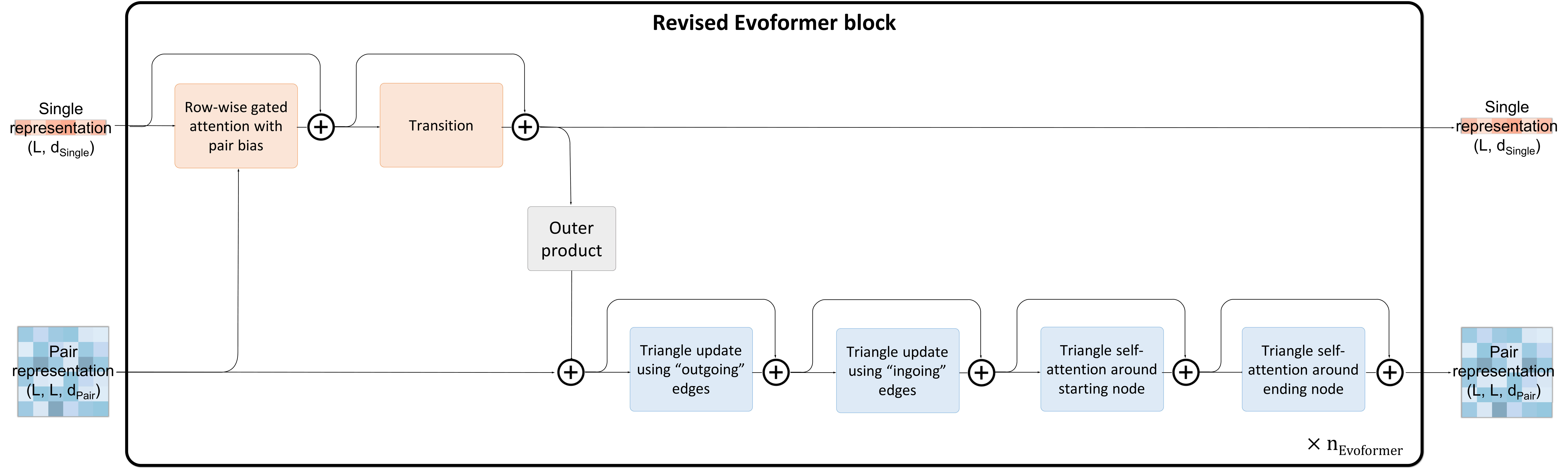}  
\caption{Architecture of revised Evoformer.}
\label{fig:evoformer}
\end{figure}

Since HelixFold-Single takes only the single sequence as input, we slightly modify the architecture of Evoformer, removing the columnwise attention. The architecture of the revised Evoformer is shown in Figure~\ref{fig:evoformer}.

\begin{table}[t]
 \caption{Median times (seconds) of MSA search, AlphaFold2, and HelixFold-Single on proteins with various lengths, detailed results corresponding to Figure~\ref{fig:efficiency}.}
  \centering
    \begin{tabular}{lccccc}
    \toprule
     & [1,100] & (100, 200] & (200, 400] & (400, 800] & (800,$+\infty$) \\ 
    \midrule
    MSA search & 737.5 & 755.4 & 853.7 & 977.0 & 1203.8 \\
    AlphaFold2 & 766.1 & 795.8 & 908.3 & 1125.2 & 1611.2 \\
    HelixFold-Single & 1.5 & 1.5 & 2.1 & 6.2 & 37.5 \\
    \bottomrule
    \end{tabular}
    \label{tab:efficiency}
\end{table}

\clearpage

\bibliographystyle{unsrt}  
\bibliography{references}

\begin{thebibliography}{10}

\bibitem{jumper2021highly}
John Jumper, Richard Evans, Alexander Pritzel, Tim Green, Michael Figurnov,
  Olaf Ronneberger, Kathryn Tunyasuvunakool, Russ Bates, Augustin
  {\v{Z}}{\'\i}dek, Anna Potapenko, et~al.
\newblock Highly accurate protein structure prediction with alphafold.
\newblock {\em Nature}, 596(7873):583--589, 2021.

\bibitem{moult2005decade}
John Moult.
\newblock A decade of casp: progress, bottlenecks and prognosis in protein
  structure prediction.
\newblock {\em Current opinion in structural biology}, 15(3):285--289, 2005.

\bibitem{vaswani2017attention}
Ashish Vaswani, Noam Shazeer, Niki Parmar, Jakob Uszkoreit, Llion Jones,
  Aidan~N Gomez, {\L}ukasz Kaiser, and Illia Polosukhin.
\newblock Attention is all you need.
\newblock {\em Advances in neural information processing systems}, 30, 2017.

\bibitem{kenton2019bert}
Jacob Devlin Ming-Wei~Chang Kenton and Lee~Kristina Toutanova.
\newblock Bert: Pre-training of deep bidirectional transformers for language
  understanding.
\newblock In {\em Proceedings of NAACL-HLT}, pages 4171--4186, 2019.

\bibitem{brown2020language}
Tom Brown, Benjamin Mann, Nick Ryder, Melanie Subbiah, Jared~D Kaplan, Prafulla
  Dhariwal, Arvind Neelakantan, Pranav Shyam, Girish Sastry, Amanda Askell,
  et~al.
\newblock Language models are few-shot learners.
\newblock {\em Advances in neural information processing systems},
  33:1877--1901, 2020.

\bibitem{rao2019evaluating}
Roshan Rao, Nicholas Bhattacharya, Neil Thomas, Yan Duan, Peter Chen, John
  Canny, Pieter Abbeel, and Yun Song.
\newblock Evaluating protein transfer learning with tape.
\newblock {\em Advances in neural information processing systems}, 32, 2019.

\bibitem{elnaggar2020prottrans}
Ahmed Elnaggar, Michael Heinzinger, Christian Dallago, Ghalia Rihawi, Yu~Wang,
  Llion Jones, Tom Gibbs, Tamas Feher, Christoph Angerer, Martin Steinegger,
  et~al.
\newblock Prottrans: towards cracking the language of life's code through
  self-supervised deep learning and high performance computing.
\newblock {\em arXiv preprint arXiv:2007.06225}, 2020.

\bibitem{rao2020transformer}
Roshan Rao, Joshua Meier, Tom Sercu, Sergey Ovchinnikov, and Alexander Rives.
\newblock Transformer protein language models are unsupervised structure
  learners.
\newblock {\em Biorxiv}, 2020.

\bibitem{xiao2021modeling}
Yijia Xiao, Jiezhong Qiu, Ziang Li, Chang-Yu Hsieh, and Jie Tang.
\newblock Modeling protein using large-scale pretrain language model.
\newblock {\em arXiv preprint arXiv:2108.07435}, 2021.

\bibitem{chowdhury2021single}
Ratul Chowdhury, Nazim Bouatta, Surojit Biswas, Charlotte Rochereau, George~M
  Church, Peter~Karl Sorger, and Mohammed~N AlQuraishi.
\newblock Single-sequence protein structure prediction using language models
  from deep learning.
\newblock {\em bioRxiv}, 2021.

\bibitem{weissenow2022protein}
Konstantin Wei{\ss}enow, Michael Heinzinger, and Burkhard Rost.
\newblock Protein language-model embeddings for fast, accurate, and
  alignment-free protein structure prediction.
\newblock {\em Structure}, 2022.

\bibitem{wang2022single}
Wenkai Wang, Zhenling Peng, and Jianyi Yang.
\newblock Single-sequence protein structure prediction using supervised
  transformer protein language models.
\newblock {\em bioRxiv}, 2022.

\bibitem{he2020deberta}
Pengcheng He, Xiaodong Liu, Jianfeng Gao, and Weizhu Chen.
\newblock Deberta: Decoding-enhanced bert with disentangled attention.
\newblock In {\em International Conference on Learning Representations}, 2020.

\bibitem{mirdita2017uniclust}
Milot Mirdita, Lars Von Den~Driesch, Clovis Galiez, Maria~J Martin, Johannes
  S{\"o}ding, and Martin Steinegger.
\newblock Uniclust databases of clustered and deeply annotated protein
  sequences and alignments.
\newblock {\em Nucleic acids research}, 45(D1):D170--D176, 2017.

\bibitem{10.1093/bioinformatics/btu739}
Baris~E. Suzek, Yuqi Wang, Hongzhan Huang, Peter~B. McGarvey, Cathy~H. Wu, and
  the UniProt~Consortium.
\newblock {UniRef clusters: a comprehensive and scalable alternative for
  improving sequence similarity searches}.
\newblock {\em Bioinformatics}, 31(6):926--932, 11 2014.

\bibitem{10.1093/nar/28.1.235}
Helen~M. Berman, John Westbrook, Zukang Feng, Gary Gilliland, T.~N. Bhat, Helge
  Weissig, Ilya~N. Shindyalov, and Philip~E. Bourne.
\newblock {The Protein Data Bank}.
\newblock {\em Nucleic Acids Research}, 28(1):235--242, 01 2000.

\bibitem{10.1093/nar/gkaa1038}
Stephen~K Burley, Charmi Bhikadiya, Chunxiao Bi, Sebastian Bittrich, Li~Chen,
  Gregg~V Crichlow, Cole~H Christie, Kenneth Dalenberg, Luigi Di~Costanzo,
  Jose~M Duarte, Shuchismita Dutta, Zukang Feng, Sai Ganesan, David~S Goodsell,
  Sutapa Ghosh, Rachel~Kramer Green, Vladimir Guranović, Dmytro Guzenko,
  Brian~P Hudson, Catherine L Lawson, Yuhe Liang, Robert Lowe, Harry Namkoong,
  Ezra Peisach, Irina Persikova, Chris Randle, Alexander Rose, Yana Rose,
  Andrej Sali, Joan Segura, Monica Sekharan, Chenghua Shao, Yi-Ping Tao, Maria
  Voigt, John D Westbrook, Jasmine~Y Young, Christine Zardecki, and Marina
  Zhuravleva.
\newblock {RCSB Protein Data Bank: powerful new tools for exploring 3D
  structures of biological macromolecules for basic and applied research and
  education in fundamental biology, biomedicine, biotechnology, bioengineering
  and energy sciences}.
\newblock {\em Nucleic Acids Research}, 49(D1):D437--D451, 11 2020.

\bibitem{10.1093/nar/gkab1061}
Mihaly Varadi, Stephen Anyango, Mandar Deshpande, Sreenath Nair, Cindy
  Natassia, Galabina Yordanova, David Yuan, Oana Stroe, Gemma Wood, Agata
  Laydon, Augustin Žídek, Tim Green, Kathryn Tunyasuvunakool, Stig Petersen,
  John Jumper, Ellen Clancy, Richard Green, Ankur Vora, Mira Lutfi, Michael
  Figurnov, Andrew Cowie, Nicole Hobbs, Pushmeet Kohli, Gerard Kleywegt, Ewan
  Birney, Demis Hassabis, and Sameer Velankar.
\newblock {AlphaFold Protein Structure Database: massively expanding the
  structural coverage of protein-sequence space with high-accuracy models}.
\newblock {\em Nucleic Acids Research}, 50(D1):D439--D444, 11 2021.

\bibitem{https://doi.org/10.1002/prot.26202}
Lisa~N. Kinch, R.~Dustin Schaeffer, Andriy Kryshtafovych, and Nick~V. Grishin.
\newblock Target classification in the 14th round of the critical assessment of
  protein structure prediction (casp14).
\newblock {\em Proteins: Structure, Function, and Bioinformatics},
  89(12):1618--1632, 2021.

\bibitem{https://doi.org/10.1002/prot.26237}
Andriy Kryshtafovych, Torsten Schwede, Maya Topf, Krzysztof Fidelis, and John
  Moult.
\newblock Critical assessment of methods of protein structure prediction
  (casp)—round xiv.
\newblock {\em Proteins: Structure, Function, and Bioinformatics},
  89(12):1607--1617, 2021.

\bibitem{https://doi.org/10.1002/prot.26213}
Xavier Robin, Juergen Haas, Rafal Gumienny, Anna Smolinski, Gerardo Tauriello,
  and Torsten Schwede.
\newblock Continuous automated model evaluation (cameo)—perspectives on the
  future of fully automated evaluation of structure prediction methods.
\newblock {\em Proteins: Structure, Function, and Bioinformatics},
  89(12):1977--1986, 2021.

\bibitem{doi:10.1126/science.abj8754}
Minkyung Baek, Frank DiMaio, Ivan Anishchenko, Justas Dauparas, Sergey
  Ovchinnikov, Gyu~Rie Lee, Jue Wang, Qian Cong, Lisa~N. Kinch, R.~Dustin
  Schaeffer, Claudia Millán, Hahnbeom Park, Carson Adams, Caleb~R. Glassman,
  Andy DeGiovanni, Jose~H. Pereira, Andria~V. Rodrigues, Alberdina~A. van Dijk,
  Ana~C. Ebrecht, Diederik~J. Opperman, Theo Sagmeister, Christoph Buhlheller,
  Tea Pavkov-Keller, Manoj~K. Rathinaswamy, Udit Dalwadi, Calvin~K. Yip,
  John~E. Burke, K.~Christopher Garcia, Nick~V. Grishin, Paul~D. Adams,
  Randy~J. Read, and David Baker.
\newblock Accurate prediction of protein structures and interactions using a
  three-track neural network.
\newblock {\em Science}, 373(6557):871--876, 2021.

\bibitem{zhang2004scoring}
Yang Zhang and Jeffrey Skolnick.
\newblock Scoring function for automated assessment of protein structure
  template quality.
\newblock {\em Proteins: Structure, Function, and Bioinformatics},
  57(4):702--710, 2004.

\bibitem{brown1992estimate}
Peter~F Brown, Stephen~A Della~Pietra, Vincent~J Della~Pietra, Jennifer~C Lai,
  and Robert~L Mercer.
\newblock An estimate of an upper bound for the entropy of english.
\newblock {\em Computational Linguistics}, 18(1):31--40, 1992.

\bibitem{rao2021msa}
Roshan~M Rao, Jason Liu, Robert Verkuil, Joshua Meier, John Canny, Pieter
  Abbeel, Tom Sercu, and Alexander Rives.
\newblock Msa transformer.
\newblock In {\em International Conference on Machine Learning}, pages
  8844--8856. PMLR, 2021.

\bibitem{radford2018improving}
Alec Radford, Karthik Narasimhan, Tim Salimans, Ilya Sutskever, et~al.
\newblock Improving language understanding by generative pre-training.
\newblock 2018.

\bibitem{yang2020improved}
Jianyi Yang, Ivan Anishchenko, Hahnbeom Park, Zhenling Peng, Sergey
  Ovchinnikov, and David Baker.
\newblock Improved protein structure prediction using predicted interresidue
  orientations.
\newblock {\em Proceedings of the National Academy of Sciences},
  117(3):1496--1503, 2020.

\bibitem{yang2015tasser}
Jianyi Yang, Renxiang Yan, Ambrish Roy, Dong Xu, Jonathan Poisson, and Yang
  Zhang.
\newblock The i-tasser suite: protein structure and function prediction.
\newblock {\em Nature methods}, 12(1):7--8, 2015.

\bibitem{du2021trrosetta}
Zongyang Du, Hong Su, Wenkai Wang, Lisha Ye, Hong Wei, Zhenling Peng, Ivan
  Anishchenko, David Baker, and Jianyi Yang.
\newblock The trrosetta server for fast and accurate protein structure
  prediction.
\newblock {\em Nature protocols}, 16(12):5634--5651, 2021.

\bibitem{peng2011raptorx}
Jian Peng and Jinbo Xu.
\newblock Raptorx: exploiting structure information for protein alignment by
  statistical inference.
\newblock {\em Proteins: Structure, Function, and Bioinformatics},
  79(S10):161--171, 2011.

\bibitem{rives2021biological}
Alexander Rives, Joshua Meier, Tom Sercu, Siddharth Goyal, Zeming Lin, Jason
  Liu, Demi Guo, Myle Ott, C~Lawrence Zitnick, Jerry Ma, et~al.
\newblock Biological structure and function emerge from scaling unsupervised
  learning to 250 million protein sequences.
\newblock {\em Proceedings of the National Academy of Sciences},
  118(15):e2016239118, 2021.

\bibitem{wang2022helixfold}
Guoxia Wang, Xiaomin Fang, Zhihua Wu, Yiqun Liu, Yang Xue, Yingfei Xiang,
  Dianhai Yu, Fan Wang, and Yanjun Ma.
\newblock Helixfold: An efficient implementation of alphafold2 using
  paddlepaddle.
\newblock {\em arXiv preprint arXiv:2207.05477}, 2022.

\bibitem{loshchilov2018decoupled}
Ilya Loshchilov and Frank Hutter.
\newblock Decoupled weight decay regularization.
\newblock In {\em International Conference on Learning Representations}, 2018.

\bibitem{kingma2014adam}
Diederik~P Kingma and Jimmy Ba.
\newblock Adam: A method for stochastic optimization.
\newblock {\em arXiv preprint arXiv:1412.6980}, 2014.

\bibitem{pascanu2013difficulty}
Razvan Pascanu, Tomas Mikolov, and Yoshua Bengio.
\newblock On the difficulty of training recurrent neural networks.
\newblock In {\em International conference on machine learning}, pages
  1310--1318. PMLR, 2013.

\bibitem{xiong2020layer}
Ruibin Xiong, Yunchang Yang, Di~He, Kai Zheng, Shuxin Zheng, Chen Xing,
  Huishuai Zhang, Yanyan Lan, Liwei Wang, and Tieyan Liu.
\newblock On layer normalization in the transformer architecture.
\newblock In {\em International Conference on Machine Learning}, pages
  10524--10533. PMLR, 2020.

\end{thebibliography}

\end{document}